\documentclass[5p]{elsarticle}

\usepackage{amssymb}
\usepackage{lineno}

\def \kpnn {K^{+} \rightarrow \pi^{+}\nu\overline{\nu}}
\makeatletter
\renewcommand*{\p@section}{\S\,}
\renewcommand*{\p@subsection}{\S\,}
\makeatother

\begin{document}
%\linenumbers

\begin{frontmatter}



\title{Development of the kaon tagging system for the NA62 experiment at CERN}

%\author{}
%
%\address{}
%
%
%***************
%   Authors
%***************
%
\author[First]{Evgueni Goudzovski\fnref{RS1,ERC1}}
\fntext[RS1]{Supported by a Royal Society University Research Fellowship (UF100308)}
\fntext[ERC1]{Supported by ERC-2013-StG Project 336581}
\author[First]{Marian Krivda}
\author[First]{Cristina Lazzeroni\fnref{RS2}}
\fntext[RS2]{Supported by a Royal Society University Research Fellowship (UF0758946)}
\author[First]{Karim Massri\fnref{liv}}
\fntext[liv]{Now at University of Liverpool}
\author[First]{Francis O. Newson}
\author[First]{Simon Pyatt}
\author[First]{Angela Romano\fnref{ERC2}}
\fntext[ERC2]{Supported by ERC-2010-AdG Project 268062}
\author[First]{Xen Serghi}
\author[First]{Antonino Sergi\corref{cor1}\fnref{STFC}}
\ead{Antonino.Sergi@cern.ch}
\cortext[cor1]{Corresponding author} 
\fntext[STFC]{Supported by a STFC Ernest Rutherford Fellowship (ST/J00412X/1)}
\author[First]{Richard J. Staley}
\author[Second]{Helen F. Heath}
\author[Second]{Ryan F. Page\fnref{ERC2}}
\author[Flo]{\\Antonio Cassese}
\author[Third]{Peter A. Cooke}
\author[Third]{John B. Dainton}
\author[Third]{John R. Fry}
\author[Third]{Liam D. J. Fulton}
\author[Third]{Emlyn Jones}
\author[Third]{\\Tim J. Jones}
\author[Third]{Kevin J. McCormick}
\author[Third]{Peter Sutcliffe}
\author[Third]{Bozydar Wrona\fnref{ERC2}}
%\author[]{}
%\author[]{}
%
%
%***************
%   Addresses
%***************
%
\address[First]     {School of Physics and Astronomy, University of Birmingham, B15 2TT, United Kingdom}
\address[Second]    {School of Physics, University of Bristol, BS8 1TL, United Kingdom}
\address[Flo]       {Dipartimento di Fisica, Universit\`{a} di Firenze, I-50125, Italy}
\address[Third]     {Department of Physics, University of Liverpool, L69 7ZE, United Kingdom}
%\address[]{}
%\address[]{}

%
%===============================================================
%
%   Abstract
%
%===============================================================
%
%
\begin{abstract}

The NA62 experiment at CERN aims to make a precision measurement of the ultra-rare decay $\kpnn$, and relies
on a differential Cherenkov detector (KTAG) to identify charged kaons at an average rate of 50 MHz in a 750 MHz unseparated hadron beam.
The experimental sensitivity of NA62 to K-decay branching ratios (BR) of $10^{-11}$ requires a time resolution for the KTAG of better than 100 ps, an efficiency better than 95\% 
and a contamination of the kaon sample that is smaller than $10^{-4}$.
A prototype version of the detector was tested in 2012, during the first NA62 technical run, in which the required resolution of 100 ps was achieved and the necessary 
functionality of the light collection system and electronics was demonstrated.

\end{abstract}

%
%
%===============================================================
%
%   Keywords
%
%===============================================================
%
%

\end{frontmatter}


%
%
%===============================================================
%
%   Main text

\section{Introduction}
The aim of the NA62 experiment \cite{p326} at CERN is a precision measurement ($10\%$) of the ultra-rare decay $\kpnn$
with a branching fraction BR=$O(10^{-10})$, that makes possible a stringent test of the Standard Model because of the small theoretical uncertainties.

In order to achieve a signal to background ratio of about 10 for $\kpnn$, 
as well as using kinematic conditions, vetoes and particle identification detectors to reject events with a BR up to 10 orders of magnitude higher than signal, NA62 will rely
on a differential Cherenkov detector (KTAG) \cite{addendum} to tag kaons within an unseparated hadron beam of about 750 MHz particles, of which kaons are about 6\%,
and reject events with interactions in the residual material of the decay volume.

The design is based on a CERN West Area CEDAR detector \cite{YRep}, a Cherenkov Differential counter with Achromatic Ring focus designed in the 1970s to discriminate kaons, pions and protons in unseparated, 
charged-particle beams extracted from the CERN SPS. The CEDAR gas volume and optics are suitable for use in NA62, but the
original photodetectors and read-out electronics are not capable of sustaining the particle rate in the NA62 beam line. Therefore a new photodetection
and read-out system has been developed to meet the NA62 requirements.
The main experimental requirements are time resolution better than 100 ps, efficiency above 95\%, contamination 
of the kaon sample below $10^{-4}$ and radiation hardness.

The NA62 CEDAR is a $\approx7$ m long tube ($\oslash\approx60$ cm) filled with $N_2$ (with an option for $H_2$) at room temperature and pressure that can be varied
from vacuum to 5 bar. Starting from the downstream end of the vessel, the internal optical system consists of a mangin mirror, a chromatic corrector, lenses and a diaphragm;
the Cherenkov light is collected,
reflected and steered onto 8 quartz exit windows (upstream end), equally spaced around the circumference of the diaphragm. The new photodetection
system collects photons exiting the quartz windows and focusses them onto spherical mirrors, which reflect them onto 256 photomultipliers (PMTs).

As a first step in the development of KTAG, in 2011 the CEDAR was equipped with its original 8 PMTs, one per quartz window distributed uniformly in azimuth. 
This setup was used to evaluate the performance of two front-end electronics options and a basic prototype of the new light collection and detection system.
The front-end electronics is based on the NINO ASIC \cite{NINO}. Two preamplifiers were tested: one new, radiation hard, design and a second, which has already been used for the NA62 RICH 
\cite{NIMARICH2007, TNSMio, NIMARICH2009}. To test the photon detection a prototype light guide was built to replace one of the original PMTs. 
A solid aluminium block was machined to produce 3 conical sections with polished surfaces that reflected the light onto 3 Hamamatsu R7400 PMTs.
The results from the test beam were complemented by Monte Carlo simulations, used to estimate the radiation dose in the experimental area (FLUKA \cite{FLUKA}) and photon 
rate and distributions (GEANT4 \cite{G4}) on the NA62 beamline. The resulting design was tested in 2012, during the NA62 technical run, with
the prototype version of KTAG.

\begin{figure}[!t]
\centering
\includegraphics[width=0.5\textwidth]{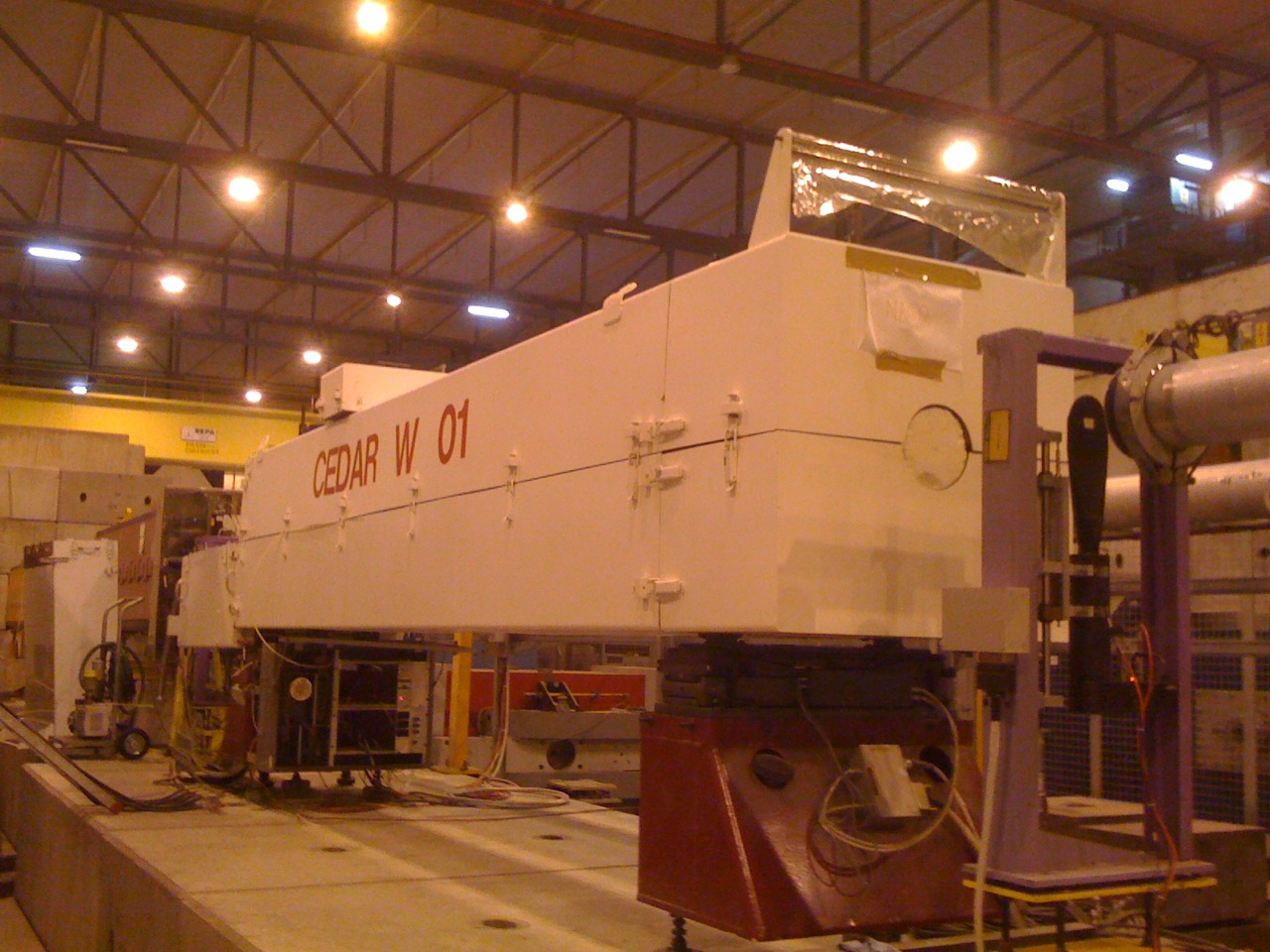}
\caption{The CEDAR installed on the H6 beam line for the 2011 test, in the North Area at CERN}
\label{fig:TB2011}
\end{figure}

\section{Test beam experimental setup}
\label{sec:TB2011}
The detector was installed on the H6 beam line at CERN in 2011 (fig. \ref{fig:TB2011}) and tested using an unseparated hadron beam of momentum 75 GeV/$c$ with a rate of $\approx40$ kHz.
The beam composition was determined as part of the test. A scintillator placed upstream of the detector was used as trigger
and beam particle counter.

\subsection{Detector}
\label{ssec:TB2011Det}
\begin{figure}[!t]
\centering
\includegraphics[width=0.5\textwidth]{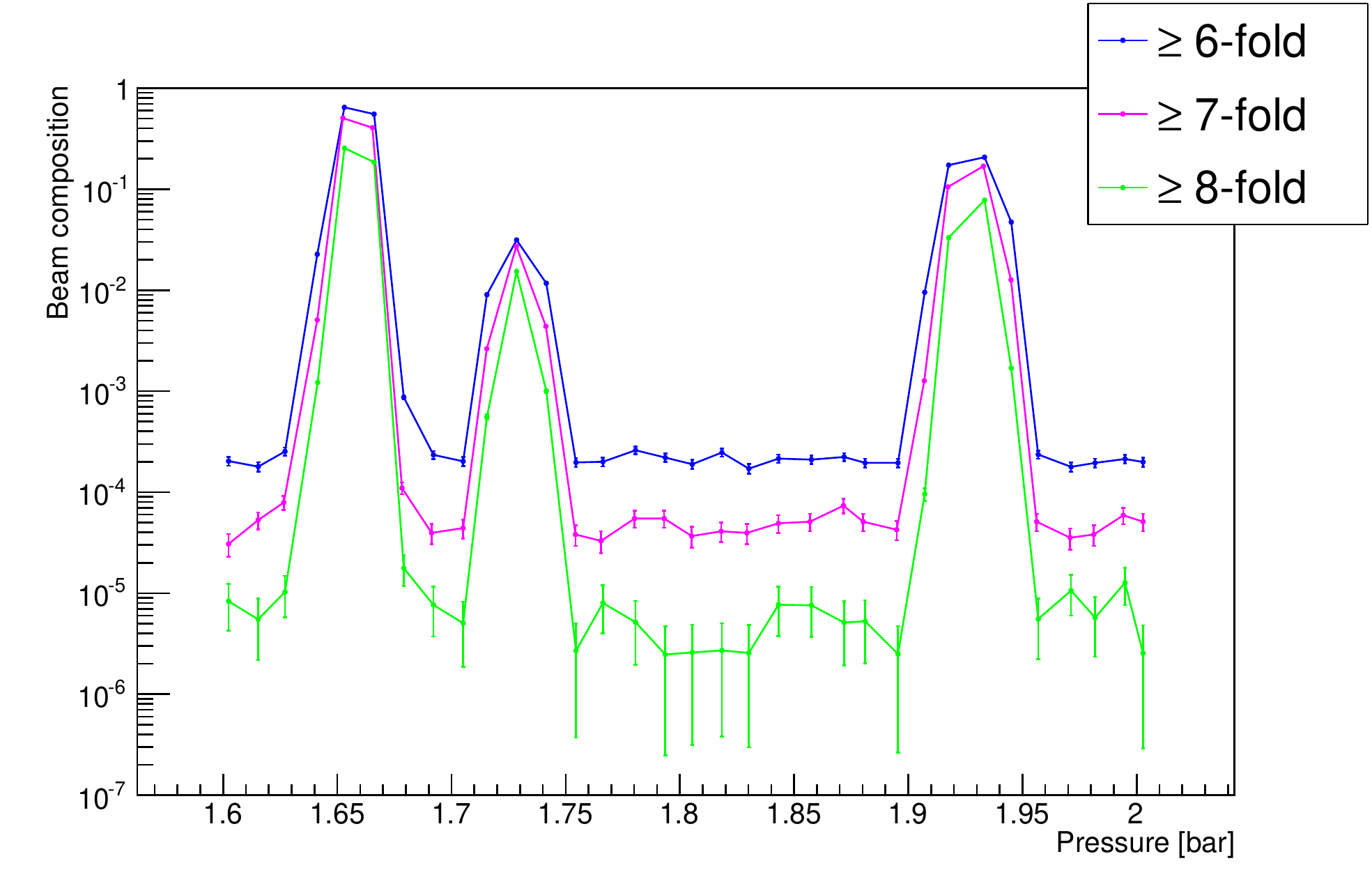}
\caption{After the alignment the pressure was varied, with a fixed diaphragm opening of $\approx1$ mm: the profile of coincidences (for at least 6, 7, and 8 PMTs) shows, in order from the left,
the pion, the kaon and the proton peaks.}
\label{fig:PScan2011}
\end{figure}

\begin{figure}[!b]
\centering
\includegraphics[width=0.5\textwidth]{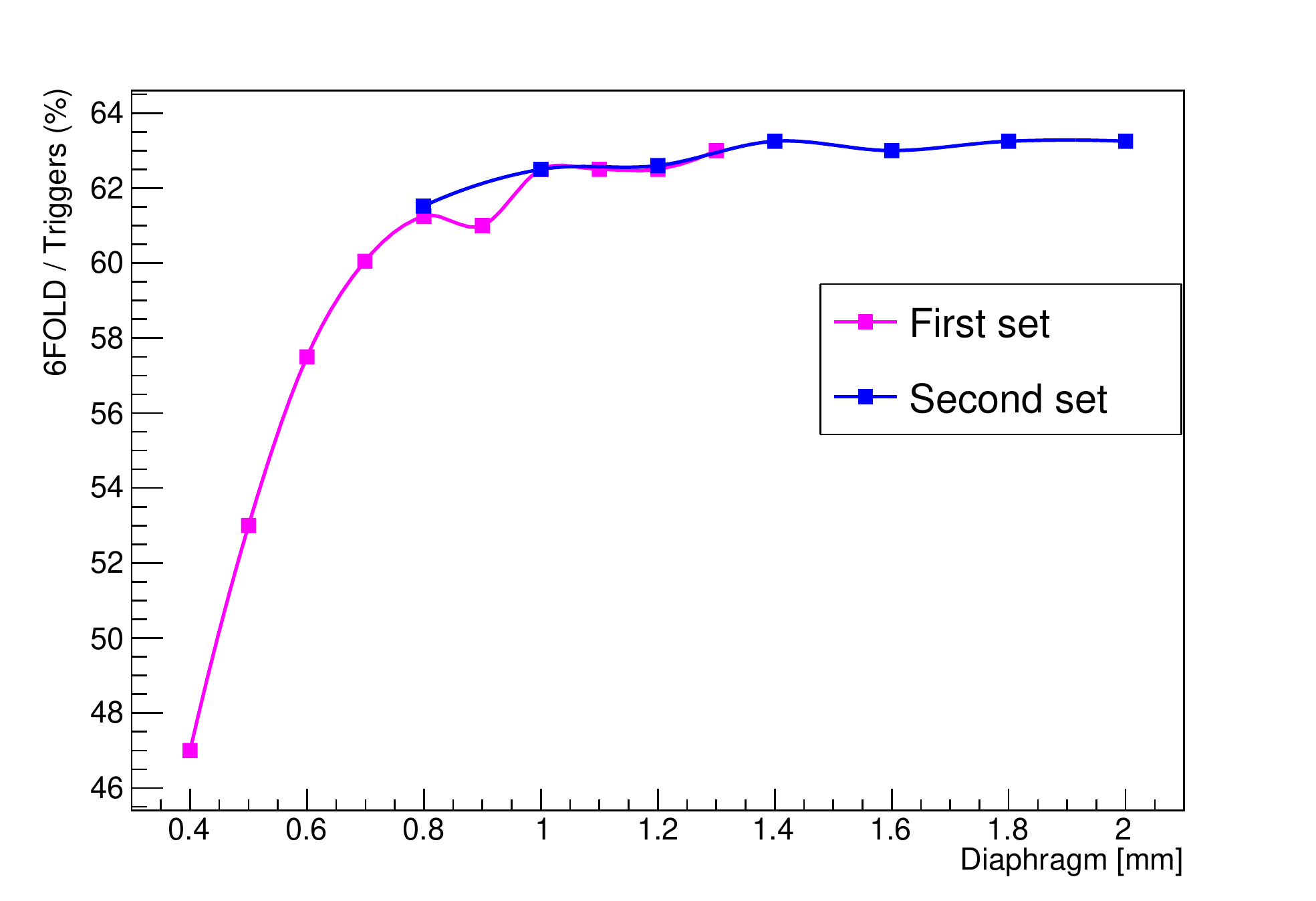}
\caption{Two data sets, at the pressure of the pion peak, showing the variation of the efficiency with the opening of the diaphragm. The statistical error is not visible on this scale.}
\label{fig:DScan2011}
\end{figure}

The CEDAR volume was filled with nitrogen at room temperature. The original read-out was used for the setup of the detector
and as a baseline for reference throughout the test. A discrimination threshold of 30mV was applied to the signals from the 8 PMTs and then
the signals were split and used for both coincidence logic and scalers. The functionality of the CEDAR optics requires incident particles to be parallel 
(to within 100 $\mu$rad) with the optical axis as they pass through the gas envelope. Alignment of the CEDAR optical axis with the beam 
is achieved by measuring the distribution of the light after it passes through an annular shaped diaphragm using the 8 PMTs. For the alignment 
a gas pressure of 1.68 bar was used, corresponding to the pion Cherenkov signal, since the pion rate is a factor 10 larger than the kaon rate. Alignment does not
require the equalisation of the count rate in the PMTs. The diaphragm aperture was gradually closed from 20 mm to its design value of 1 mm while 
adjusting the orientation of the CEDAR optical axis horizontally and vertically. Alignment is achieved when the ratios between the count rates in 
the PMTs remain constant while varying the diaphragm aperture. Finally, to confirm the functionality of CEDAR following the alignment 
procedure, the overall PMT rate was measured as a function of nitrogen pressure. Figure \ref{fig:PScan2011} demonstrates discrimination 
of pions, kaons and protons, and establishes that kaons can be cleanly selected. Figure \ref{fig:DScan2011} demonstrates that there was 
no loss of signal for diaphragm apertures of 1 mm and above. 

Once the tuning was complete, the analog signals from the PMTs were split and fed into the prototype electronics, after a 32 dB attenuation.

\subsection{Electronics}
\begin{figure}[!t]
\centering
\includegraphics[trim=3.5cm 8.5cm 4cm 9cm, clip=true,width=0.5\textwidth]{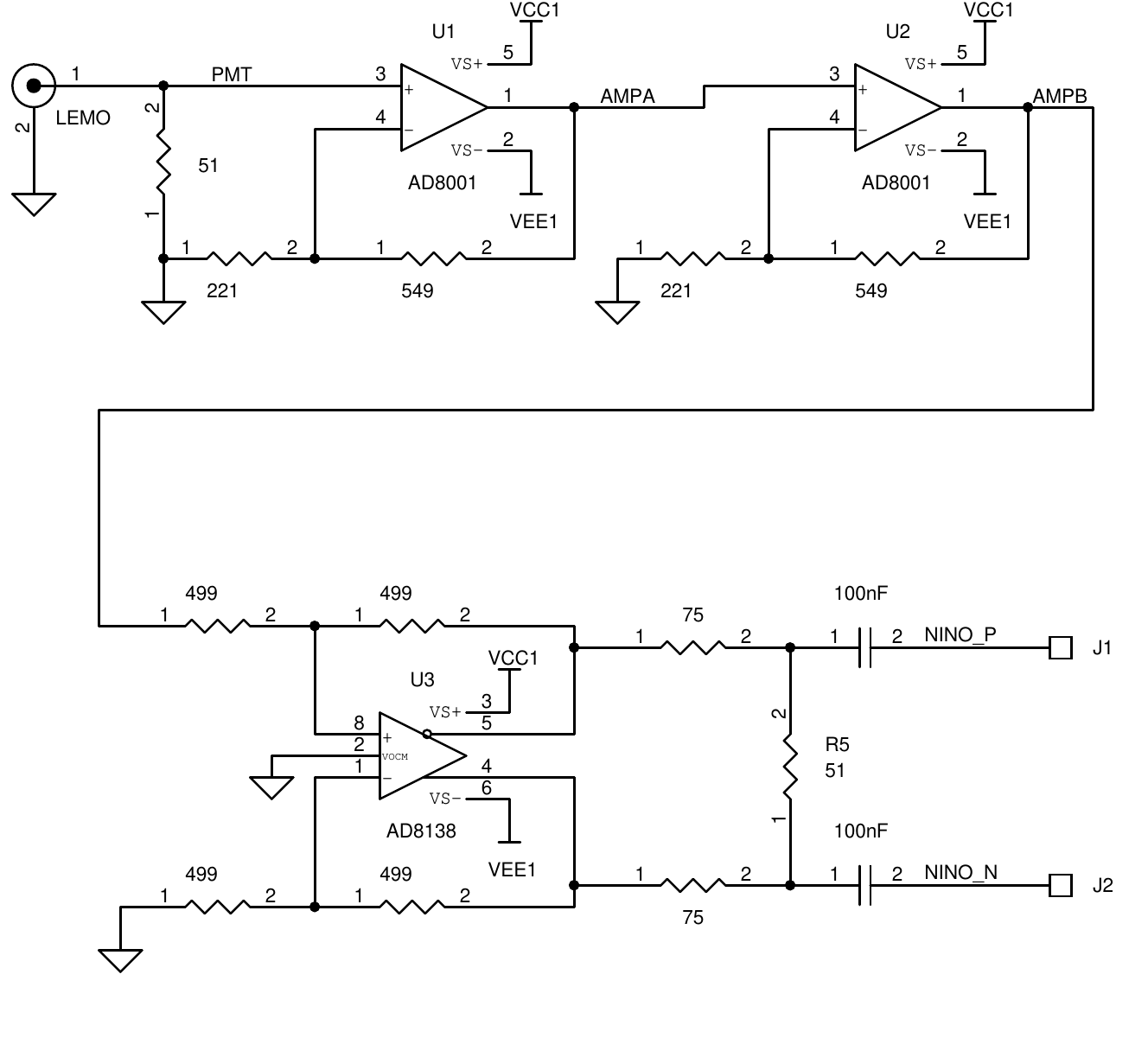}
\caption{Schematic of the radiation hard preamplifier}
\label{fig:RardHardSch}
\end{figure}

\begin{figure}[!b]
\centering
\includegraphics[width=0.5\textwidth]{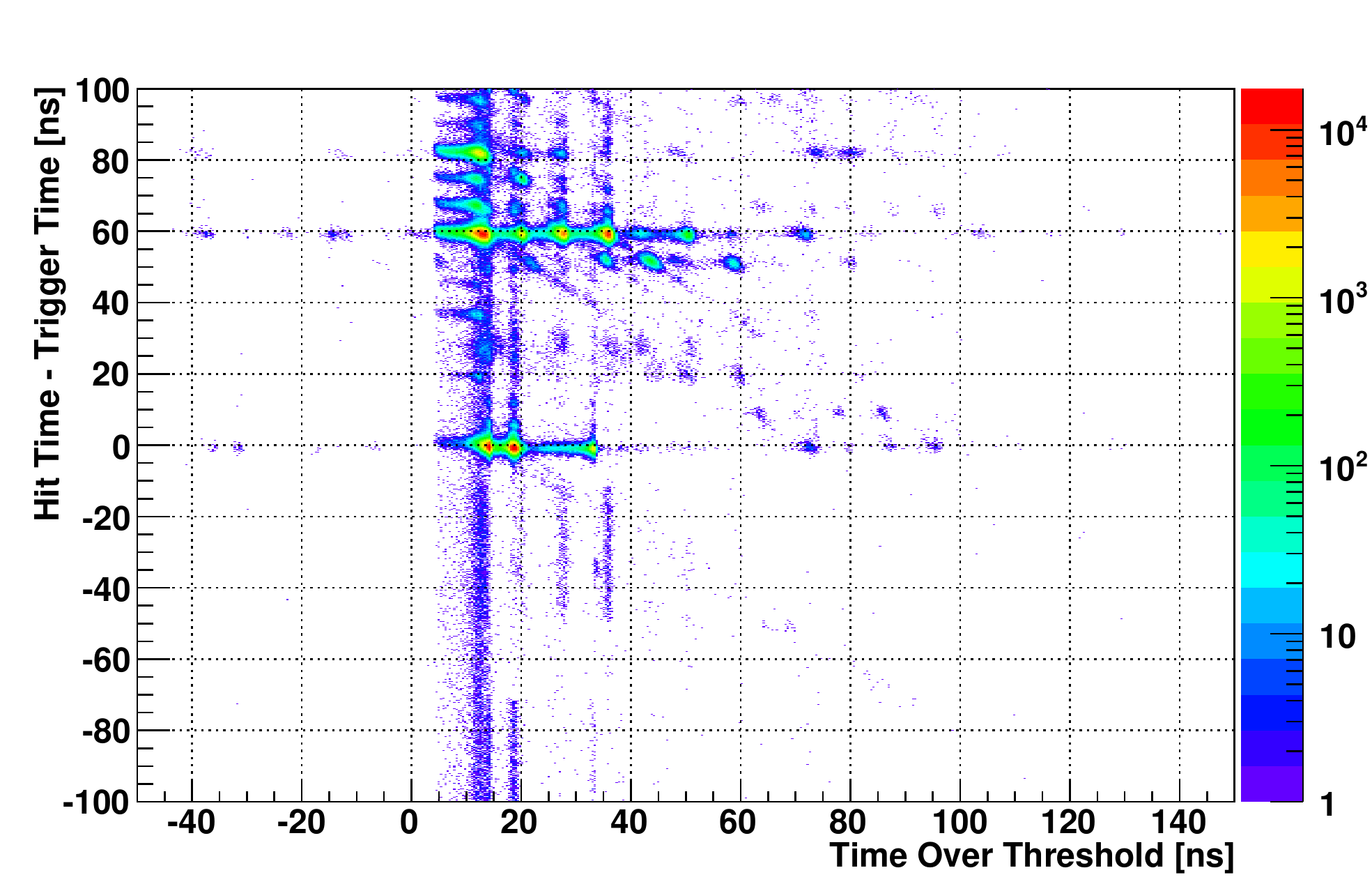}
\caption{Time performance of the radiation hard preamplifier: there is much more uncorrelated noise with respect to fig. \ref{fig:RICHT}, and the overall response suggests a distortion of the signal.}
\label{fig:RadHardT}
\end{figure}

The front-end electronics was based on NINO ASIC \cite{NINO}, a time over threshold charge discriminator with very low jitter ($\approx60$ ps). The
time over threshold technique allows an off-line slewing correction for better time resolution.
The NINO has
a differential input, for which two prototype preamplifiers were used. The newly designed (fig. \ref{fig:RardHardSch}) and radiation hard preamplifier was tested for noise and time
performance (fig. \ref{fig:RadHardT}). The second preamplifier had already been characterized for
the NA62 RICH detector and was used as a reference and also in the second stage of the tests which
studied the new PMTs.
%Fig. \ref{fig:RardHardSch} shows the schematics of the rad-hard preamplifier. 

The read-out system was based on HPTDC ASIC \cite{HPTDC} and TELL1 \cite{TELL1}; the components were part of an intermediate development stage of the final common NA62 read-out
system, composed of custom HPTDC based daughter boards and TEL62 \cite{TDAQ}.

%\begin{figure}[!t]
%\begin{minipage}{0.46\textwidth}
%\centering
%\includegraphics[width=0.4\textwidth]{RadHardT}
%\caption{Time performance of the radiation hard preamplifier: there is much more uncorrelated noise with respect to fig. \ref{fig:RICHT}, and the overall response suggests a distortion of the signal.}
%\label{fig:RadHardT}
%\end{minipage}\hfill
%\begin{minipage}{0.46\textwidth}
%\centering
%\includegraphics[width=0.4\textwidth]{RICHT}
%\caption{Time performance of the RICH preamplifier: secondary peaks along the vertical axis ($>0$) are due to signal reflections introduced by the attenuators.}
%\label{fig:RICHT}
%\end{minipage}\hfill
%\end{figure}

\begin{figure}[!t]
\centering
\includegraphics[width=0.5\textwidth]{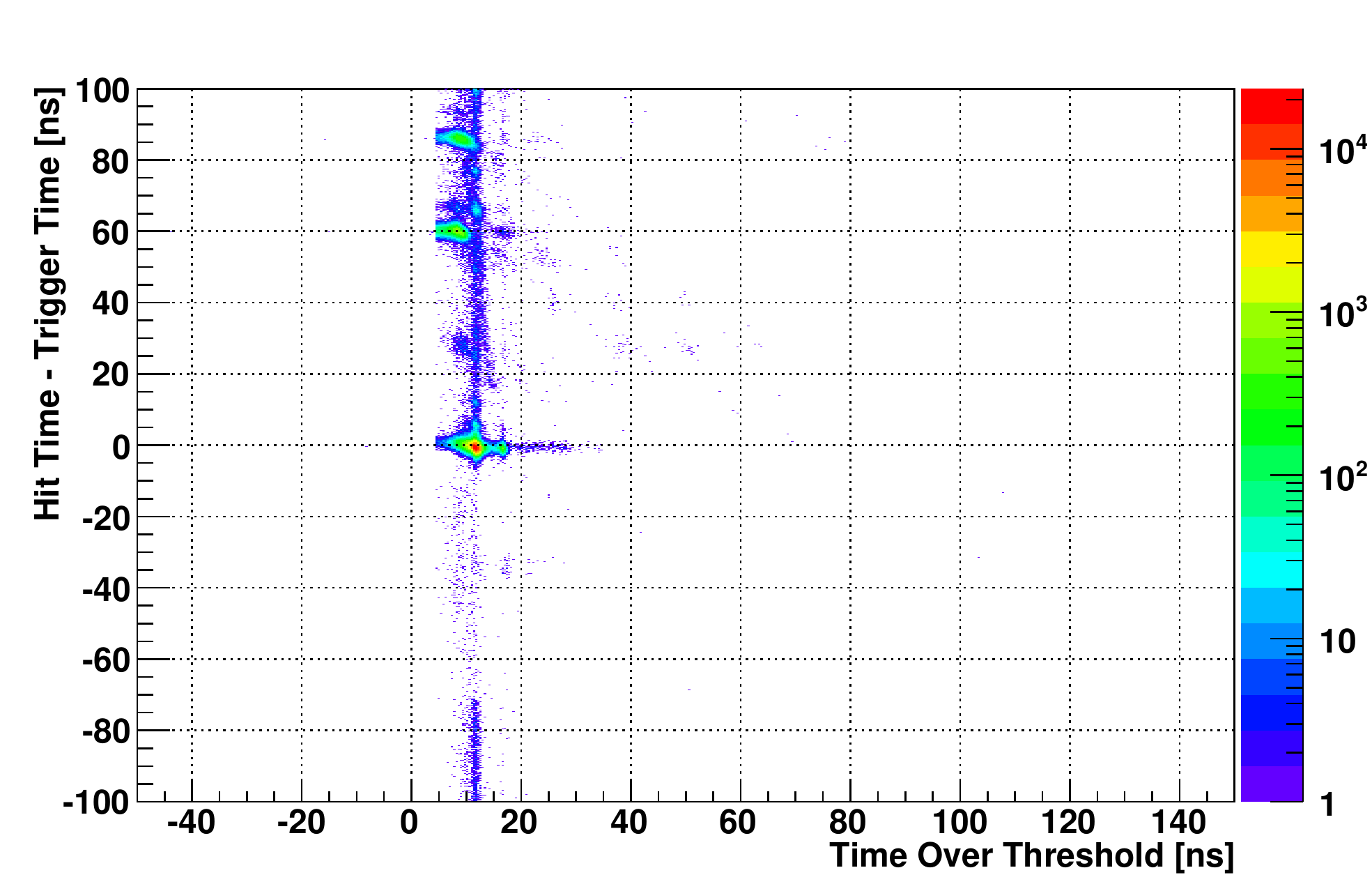}
\caption{Time performance of the RICH preamplifier: secondary peaks along the vertical axis ($>0$) are due to signal reflections introduced by the attenuators.}
\label{fig:RICHT}
\end{figure}

\subsection{Photomultiplier Tubes}
\begin{figure}[!b]
\centering
\includegraphics[trim=2.5cm 9.5cm 1.5cm 7cm, clip=true,width=0.43\textwidth]{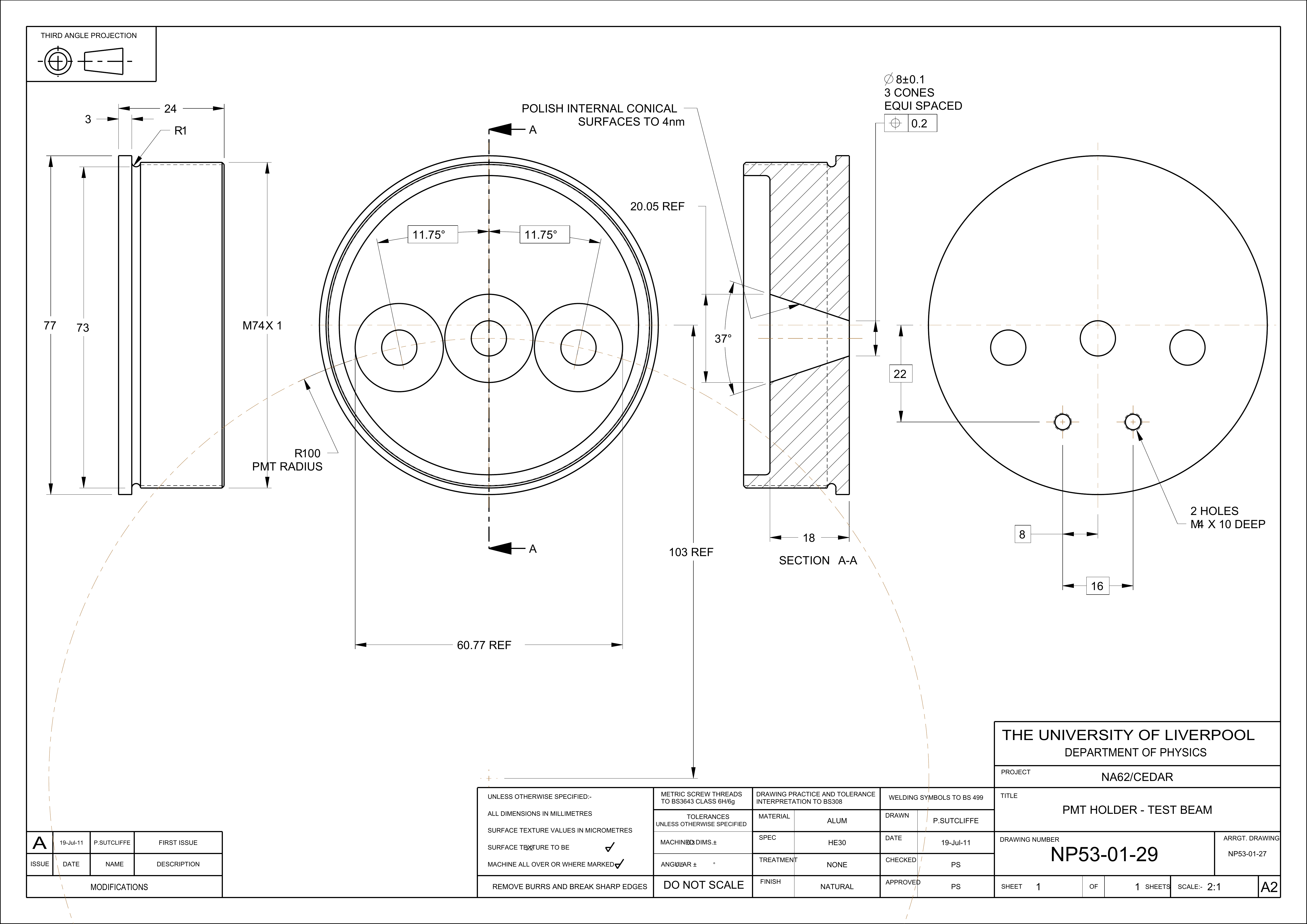}
\caption{Schematic drawing of the 3 PMTs prototype used in the 2011 test beam.}
\label{fig:PMSpotProto}
\end{figure}

\begin{figure}[!t]
\centering
\includegraphics[width=0.45\textwidth]{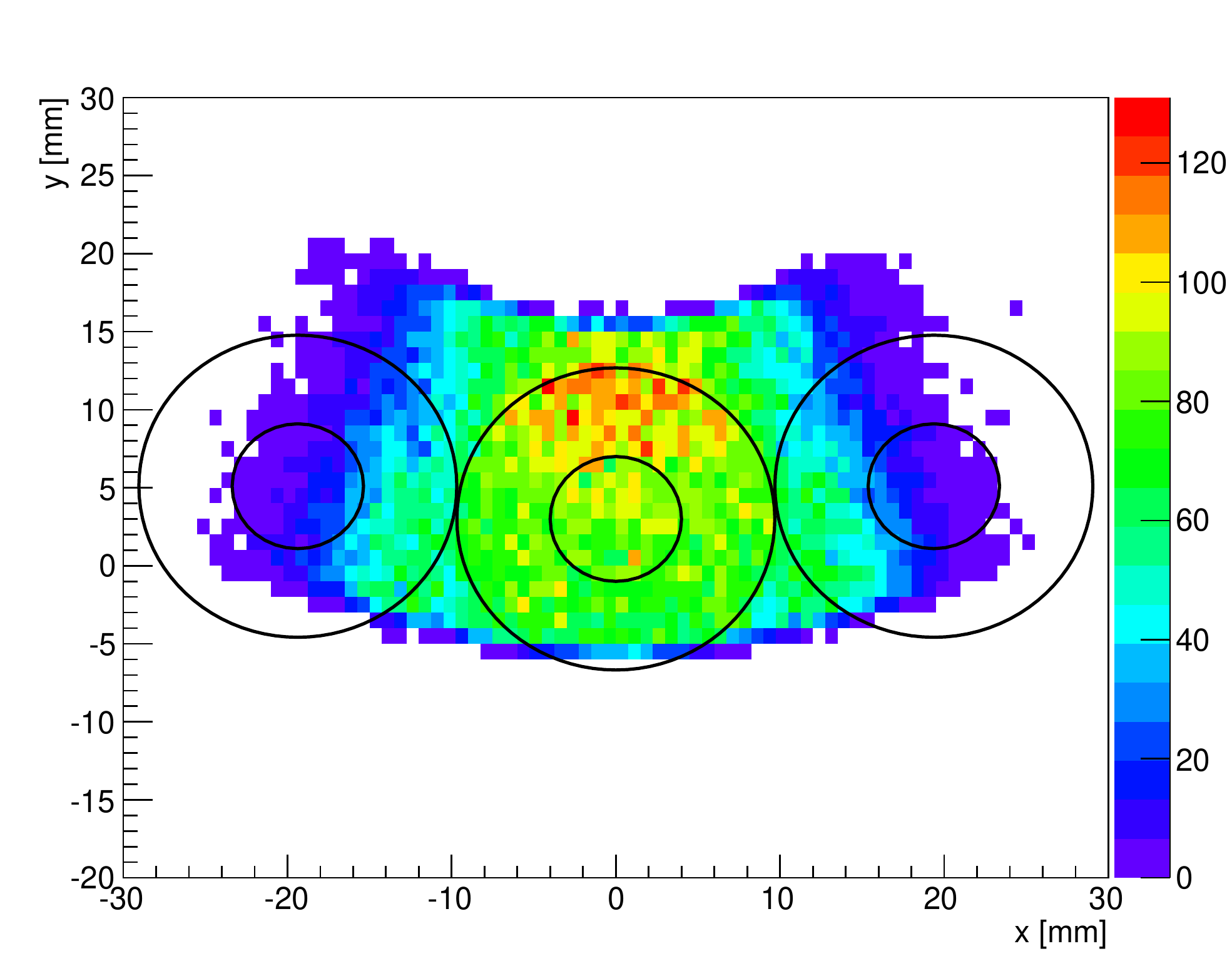}
\caption{Simulation of the illumination of the 3 PMTs prototype, in a specific configuration, with cones projection overlaid.}
\label{fig:PMSpotProtoIll}
\end{figure}

As already mentioned, the original light collection and detection system is not suitable for the intensity of NA62 beam, that will produce a photon flux of
few MHz/mm$^2$ at the CEDAR exit windows. It was therefore decideded to replace the 8 PMTs by 8 groups of small, fast Hamamatsu metal package PMTs of type R7400-U03.
Since the active area of the PMTs is $<$20\% of their geometrical area, light guides are necessary to channel the Cherenkov photons. 
%In the foreseen design for each group of PMTs the light guide allows to arrange the PMTs on a portion of a spherical surface, to minimize the correlation 
%between the incident angle of an optical photon and its impinging position on the array. In order to maximize the sensitive area (fig. \ref{fig:PMSpot}), on which the
%light is spread by a convex spherical mirror, each PMT is preceded, in the optical path, by a cone.
A prototype light guide was made, consisting of three conic sections hollowed out of an aluminium plate (fig. \ref{fig:PMSpotProto}). 
The faces of the cone were highly polished and their reflectivities measured in the laboratory. This prototype was designed to replace precisely one of the original 
CEDAR PMTs in such a way that its orientation could be varied in addition to its distance from the CEDAR quartz exit window. In this way it was possible to take data 
with different geometrical configurations in order to enable different comparisons with the Monte Carlo simulation of the light envelope produced at the prototype by 
the Cherenkov photons (fig. \ref{fig:PMSpotProtoIll}). 

\begin{figure}[!b]
\centering
\includegraphics[width=0.45\textwidth]{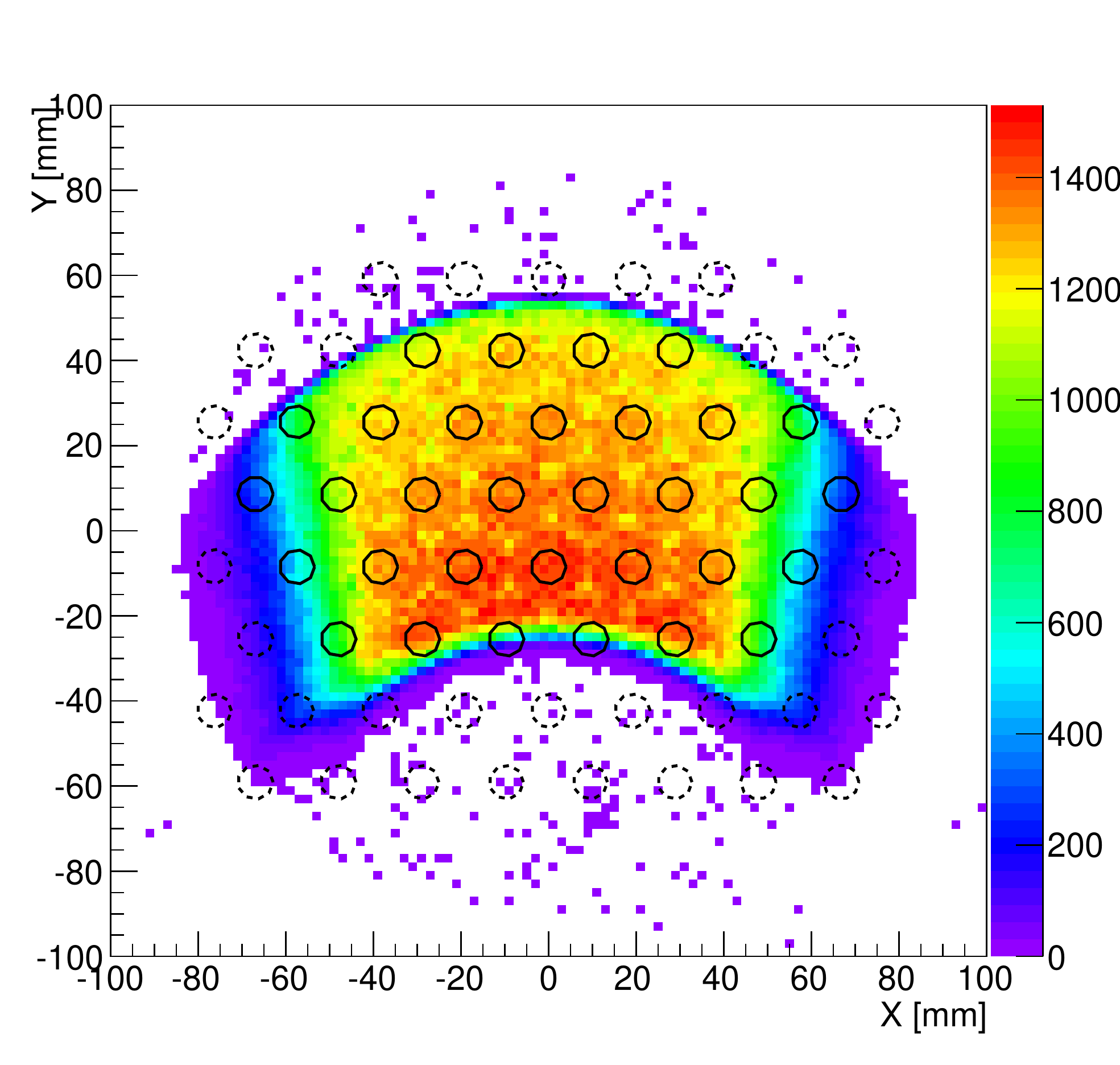}
\caption{Distribution of optical photons at the entrance plane of the cones. The array of PMTs is shown; the dashed ones are not installed.}
\label{fig:PMSpot}
\end{figure}

\section{Finalizing the design of KTAG}
Simulations with FLUKA indicate that the radiation flux at the front end
electronics is up to 0.4Gy/year and that radiation hardness is an important requirement. The
radiation hard preamplifier did not perform
as expected during the test beam. Figs. \ref{fig:RadHardT} and \ref{fig:RICHT}
show the time with respect to the trigger and time over threshold distributions for the two preamplifiers. The radiation hard preamplifier
proved to be too noisy for this application.

To remove the need for a preamplifier a new voltage divider with differential
output was designed to feed the PMT signal directly into the NINO input. This approach was expected to
introduce several percent inefficiency on some PMTs but has several intrinsic advantages: avoiding the introduction of any further noise source, partially compensating 
for the lack of a preamplifier with a gain factor of $\approx$ 2 because of the differential output, and reducing sensitivity to
common mode noise.

Data collected during the test beam enabled a realistic estimate to be made of the number of photoelectrons that would be generated by the photon flux produced by the nominal NA62 beam intensity. 
This was crucial in refining the Monte Carlo simulation describing the evolution of the Cherenkov photon flux through the optical system, and enabled the optimisation of the optical components 
(lenses, mirrors, light guides) and the number and distribution of PMTs. Fig \ref{fig:PMSpot} shows an example of the distribution of Cherenkov photons incident upon the lightguide in one octant for one such optimisation.
These studies revealed severe constraints on the rate capability of the read-out electronics.

\subsection{Monte Carlo simulation}
\begin{figure}[!t]
\centering
\includegraphics[width=0.5\textwidth]{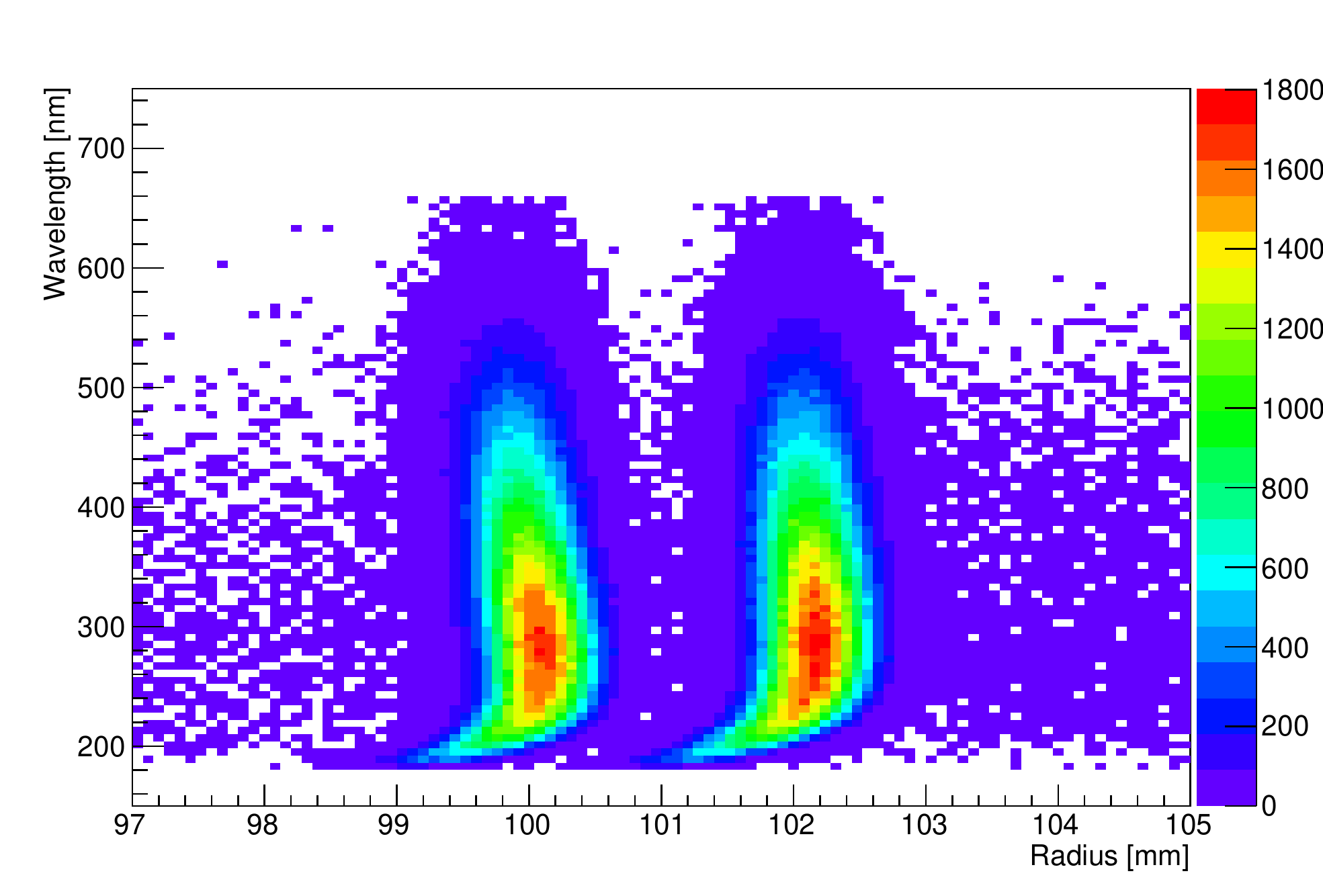}
\caption{Diaphragm illumination for $N_2$, weighted with quantum efficiency. The left hand side distribution is due to kaons.}
\label{fig:DiaphIllumN2}
\end{figure}

\begin{figure}[!t]
\centering
\includegraphics[width=0.5\textwidth]{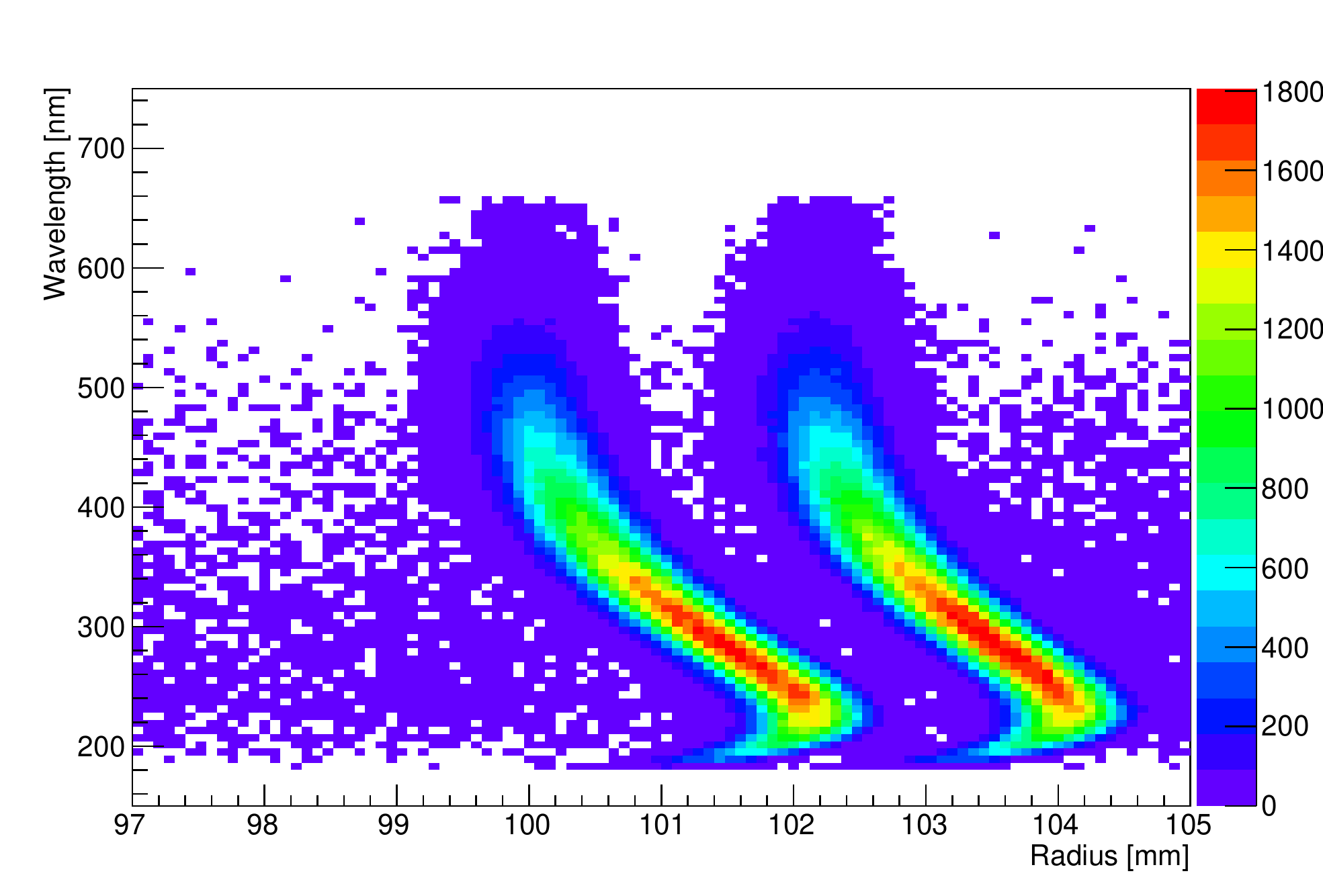}
\caption{Diaphragm illumination for $H_2$, weighted with quantum efficiency. The left hand side distribution is due to kaons.}
\label{fig:DiaphIllumH2}
\end{figure}

A full simulation of the CEDAR was undertaken in which Cherenkov photons were created within the gaseous radiator and traversed through the optical 
system to exit through the diaphragm aperture and out from the 8 quartz windows. Detailed information concerning the reflectivity and absorption 
of the different optical components as a function of wavelength was used together with the dispersive properties of the two candidate radiators, 
gaseous nitrogen and hydrogen, and the spectral response of the PMTs. The simulation shows that an average of 18 hits per beam particle is enough
to fulfil all the requirements. The West Area CEDAR used by NA62 was designed to operate with nitrogen, 
where full compensation of the dispersion was incorporated into the optics. Figure \ref{fig:DiaphIllumN2} shows that sufficient angular separation of 
kaons and pions, plotted as their radius on the diaphragm plane, is preserved at all wavelengths. The dispersion is not fully corrected with hydrogen 
and light of different wavelengths is spread over an angular range of size similar to the separation between kaons and pions (fig. \ref{fig:DiaphIllumH2}). 
%It may still be possible to choose the combination of gas pressure and a smaller diaphragm aperture to select charged kaons and fulfil the requirement 
%of a pion contamination of no more than $10^{-4}$ if the CEDAR can be aligned very precisely with the incident charged-particle beam, but this will 
%involve a loss of more than 40\% of the Cherenkov light from kaons and consequent, but affordable, degradation in the timing response for KTAG.
The design pion contamination requirement may be achievable with hydrogen at the expense of the loss of about 40\% of the Cherenkov light from kaons.
The corresponding KTAG time resolution is within design specification, while the degradation in the kaon identification efficiency would be
potentially critical.
The KTAG has been designed to be compatible with both solutions, because nitrogen offers a better optical performance and hydrogen introduces less material 
on the NA62 beam line.

The number of detectable photoelectrons per beam particle determines the efficiency, the time resolution and, to some extent, the contamination of pions
in a kaon sample. Critically for NA62, it also determines the single channel hit rate, which imposes constraints on the read-out electronics.
It is therefore necessary, within the optimization process, to prevent individual channels from exceeding 5 MHz average rate (see \ref{ssec:RO}).
Based on the test-beam measurements and studies described in \ref{sec:TB2011}, the simulated performance of KTAG for a sample of PMT configurations 
within this limit, each requiring different combinations of lenses and spherical mirrors, is shown in table \ref{tab:MCAllPerf} for both gases that might be used as the Cherenkov radiator. 
The requirements of kaon tagging efficiency greater than 95\%, based on the coincidence of signals from one or more PMTs in at least 6 octants, with an average production rate of 
photo-electrons no more than 5 MHz per PMT can be achieved for both nitrogen and hydrogen, while sufficient flexibility exists to enable further refinement of the type and configuration 
of PMTs to be used in the final design to improve the balance between kaon-tagging efficiency and rate per PMT.
For the hydrogen option PMTs with higher quantum efficiency (R9880-U110) are required.

\begin{table}[!t]
\begin{center}
\begin{tabular}{|c|c|c|c|c|}
\hline
option          &       R[mm]  &   $<$N$>$     &       MR [MHz]&       $\epsilon (\ge6)$   \\
\hline
$N_2$ R7400-U03 &       51.68  &       11.1    &       2.7       &        69\%              \\
$N_2$ R7400-U03 &       77.52  &       17.4    &       4.8       &        95\%              \\
$H_2$ R7400-U03 &       51.68  &        8.1    &       1.9       &        48\%              \\
$H_2$ R7400-U03 &       77.52  &       12.5    &       3.4       &        80\%              \\
$H_2$ R9880-U110&       77.52  &       22.4    &       5.8       &        99\%              \\
\hline
\end{tabular}
\caption{Monte Carlo estimated performance for several configurations: $N_2$ and $H_2$ options with different spherical mirror radii R and different PMT models.
The average number of hits $<$N$>$ per beam particle, the hit rate on the most active channel rate MR and the CEDAR efficiency $\epsilon$ for $\ge6$ fold coincidece is shown.}
\label{tab:MCAllPerf}
\end{center}
\end{table}

The pion contamination is dominated by accidental coincidences with kaons, hence it is determined
by the combined time resolution of the KTAG and the RICH; for a rate of charged pions in the beam of $\approx$500 MHz the requirement for both detectors is 100 ps resolution. 

\subsection{Mechanics and Optics}
To handle an instantaneous kaon flux of $\approx$50 MHz within an unseparated beam of $\approx$750 MHz, Monte Carlo simulations indicated the need for several hundred Hamamatsu PMTs 
distributed equally among the octants, with each octant replacing one of the original CEDAR PMTs and accepting light from a single quartz window. The physical size of each group of 
PMTs required that the octants were placed around the beam pipe, with light reflected radially out towards them. The radial location of the PMTs was chosen to minimise the expected 
radiation from neutrons and muons, simulated with FLUKA, and is between 30 and 50 cm from the beam line. The Cherenkov light emerging through a quartz window displays different 
angular behaviour in the radial and transverse directions and it was first thought that an ellipsoidal mirror would be required to focus the reflected light onto any sensible 
grouping of PMTs in an octant. However, detailed calculations showed that a combination of a spherical lens covering each quartz window and a spherical mirror reflecting light 
radially outwards would be satisfactory. This option greatly simplified and speeded up the fabrication, since eight pairs of standard lenses could be bought, with one from each 
pair converted to a mirror of high reflectivity at all wavelengths using an aluminium coating procedure developed by the CERN optics group.

\begin{figure}[!t]
\centering
\includegraphics[page=2,trim=56.5cm 7cm 42cm 61cm, clip=true,width=0.4\textwidth]{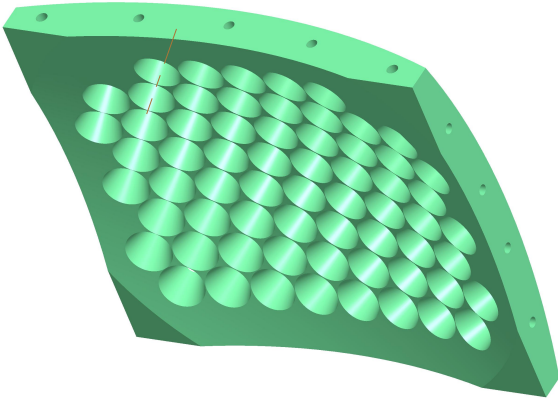}
\caption{Aluminum support for PMTs (up to 64), as a portion of a spherical shell, with cones integrated in the design.}
\label{fig:LightGuideDrawing}
\end{figure}

\begin{figure}[!b]
\centering
\includegraphics[width=0.4\textwidth]{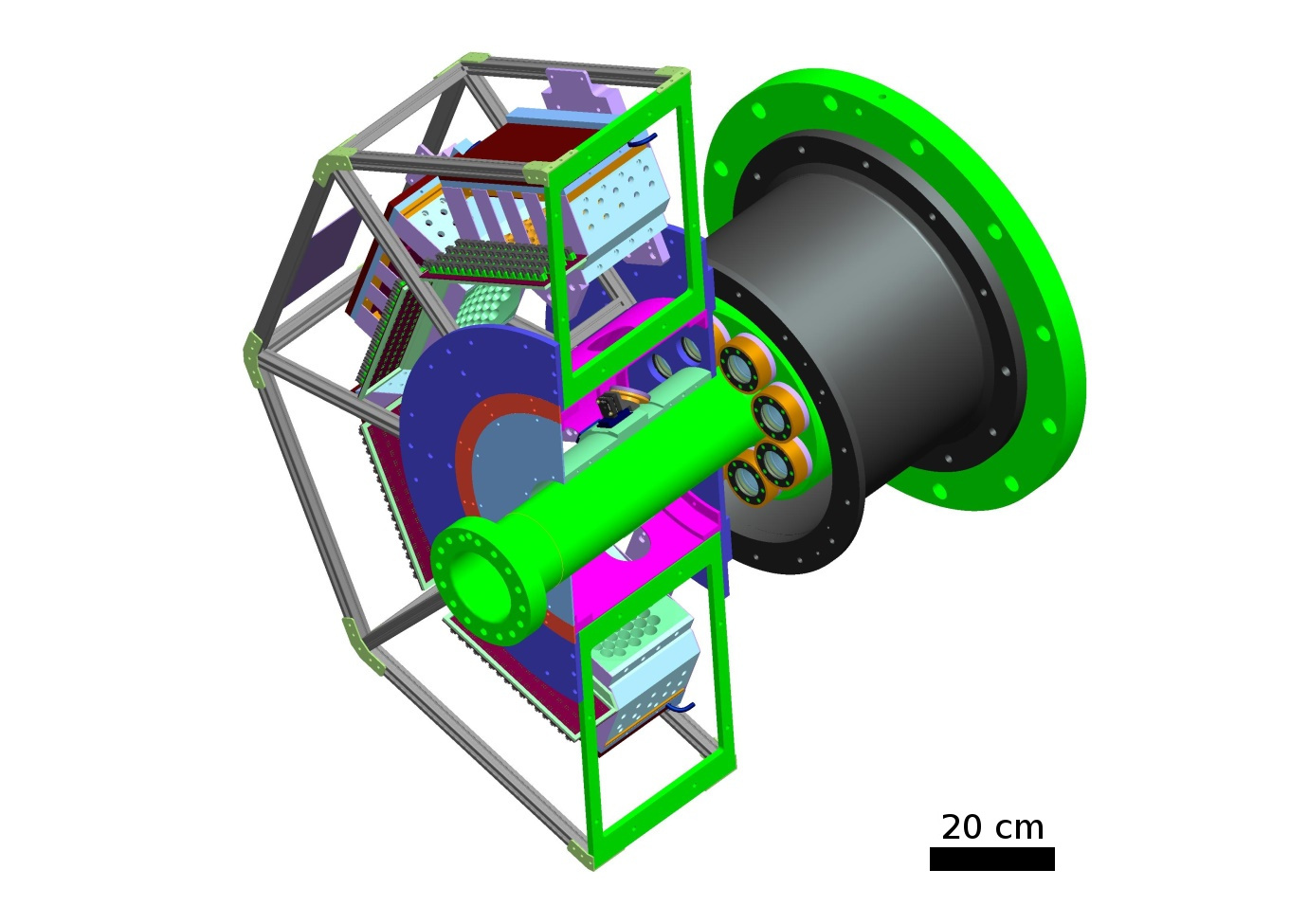}
\caption{Mechanical design of the KTAG detector, constructed in two halves to enable installation around the beam pipe. Each half comprises 4 octants and is bolted to a support 
cylinder cantilevered from the end flange of the CEDAR.}
\label{fig:KTAGMechanics}
\end{figure}

The light guides for each octant consist of a matrix of closely-spaced, conic-sections cut into an aluminium plate, with the interior of each cone lined with aluminised Mylar. 
The coating was developed by the CERN optics group to ensure high reflectivity over all relevant wavelengths. Little flexibility existed in the choice of conic parameters and in 
order to maximise the fraction of incident light reflected from the sides of the cones onto the active centre of the PMTs, the aluminium plate was machined to form a spherical 
surface with the axes of the cones converging at the virtual source of light reflected in the spherical mirror. The PMTs are set into the outer curved surface of the light guide 
to mate precisely with the cones. The light guides were made in the Liverpool University workshop to accommodate 64 PMTs (fig. \ref{fig:LightGuideDrawing}), since this was the maximum number 
foreseen on grounds of cost and performance.

KTAG is constructed in two halves to enable installation around the beam pipe, with each half comprising 4 octants (fig. \ref{fig:KTAGMechanics}). Cherenkov light exiting a quartz window is 
focussed onto a spherical mirror and reflected radially outwards onto a light guide, which forms the inner wall of a closed Light Box (LB) acting as a Faraday cage to contain the PMTs and 
readout electronics. A cooled, aluminium, heat-sink forms the outer wall and is in thermal contact with the NINO card that generates most of the heat. A cylinder surrounding the beam pipe 
holds the spherical mirrors, mounted such that their positions can be adjusted both radially and along the direction of the beam to accommodate mirrors of different radius and thickness. 
Between the mirrors and LBs is a lightweight aluminium cylinder (purple in fig. \ref{fig:KTAGMechanics}) with apertures to allow the passage of light. Both surfaces of the cylinder are matt black to 
absorb any scattered light and prevent optical cross talk between the octants. A blue LED, mounted externally to KTAG, feeds a set of optical fibres, with each fibre directing light onto 
one of the spherical mirrors. The light intensity can be adjusted using neutral-density filters and by varying the electrical current fed to the LED, and the system enables the functionality 
of the optical and electronics chains to be tested prior to the arrival of beam particles.

\begin{figure}[!t]
\centering
\includegraphics[width=0.4\textwidth]{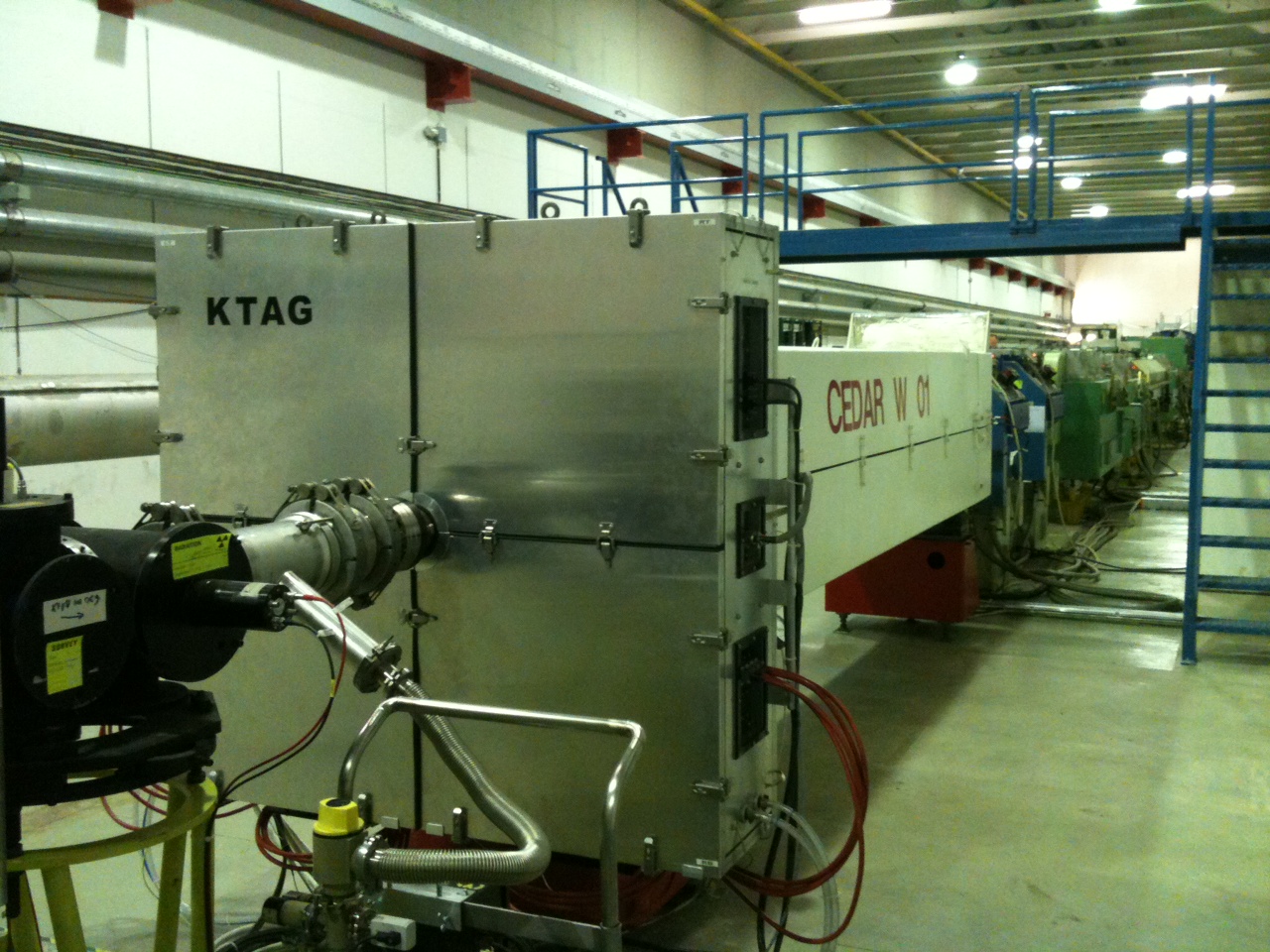}
\caption{The environmental chamber enclosing KTAG is shown in its final position on the beam line as part of the NA62 experiment in the North Area at CERN. The CEDAR gas enclosure can be seen behind KTAG.}
\label{fig:KTAGPicture}
\end{figure}

The two halves of KTAG are bolted onto a support cylinder, shown in black in fig. \ref{fig:KTAGMechanics}, which is cantilevered off the CEDAR end flange, shown in green. The complete system is enclosed 
within a light-tight, aluminium, environmental chamber (fig. \ref{fig:KTAGPicture}) lined with fire-resistant, insulating foam and continuously flushed with nitrogen gas. The nitrogen ensures that all optical components 
are kept free from dust and oxidation and also that any hydrogen that might leak through the seals on the quartz windows from the CEDAR volume is diluted and removed before any hazardous build up 
can occur. Distilled water from a chiller is fed under pressure through two sets of stainless-steel pipework, one circuit for each half of the detector, passing through the heat-sinks on the outside 
of each LB. The temperature of the water is controlled to $\pm0.1^\circ$C and the system was designed to limit the temperature drop between input and output to less than 0.5$^\circ$C. Additional 
fire-resistant insulation covers the support cylinder and other exposed regions between KTAG and CEDAR, while the CEDAR gas volume is enclosed in a well-insulated cylinder. Prior laboratory studies 
have shown that the combined effects of thermal insulation, together with the temperature-controlled KTAG environment, are sufficient to make negligible any temperature fluctuations in the CEDAR 
gas that would otherwise cause changes in refractive index and broaden the Cherenkov cone. CEDAR is equipped with thermocouples to measure the temperature at the centre and both ends of the gas 
volume, while there are 12 thermocouples distributed throughout the KTAG enclosure to monitor temperatures close to the beam pipe, the support tube, and the NINO cards. All thermocouples are read 
out and the temperatures available in real time as part of the detector monitoring.

\subsection{Front-end electronics}

\begin{figure}[!b]
\centering
\includegraphics[width=0.42\textwidth]{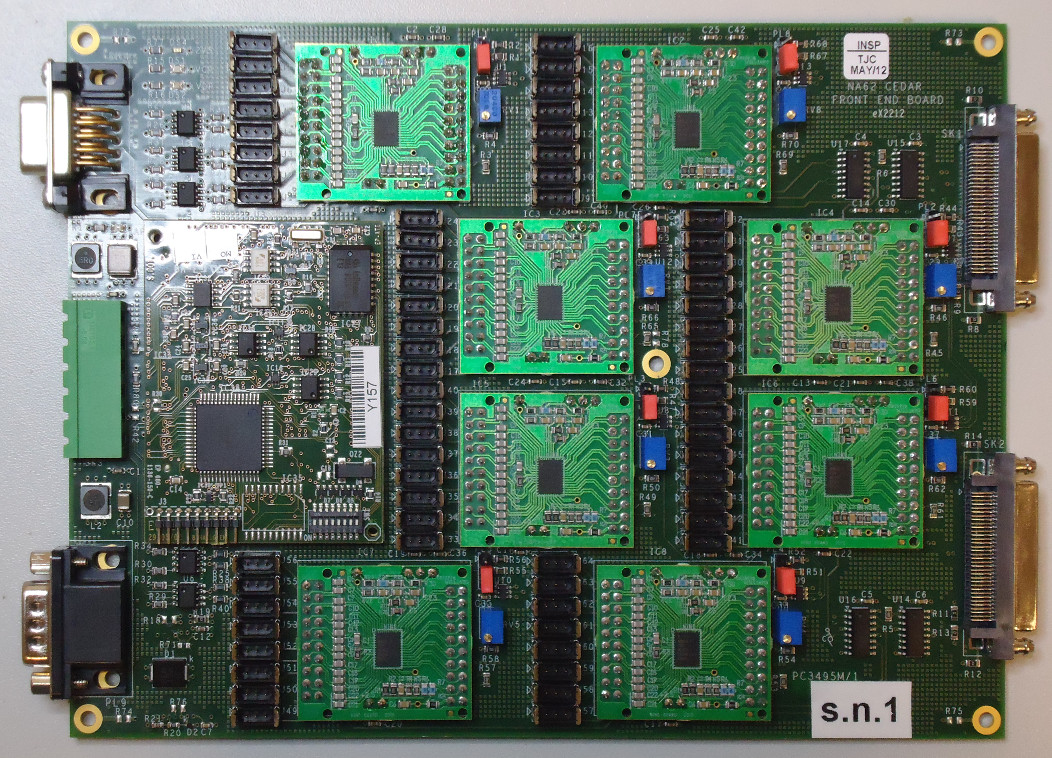}
\caption{Prototype front-end board for 2012 run, equipped with 8 NINO mezzanines.}
\label{fig:NINOBoard}
\end{figure}

To eliminate the need for a preamplifier the standard Hamamatsu voltage divider was replaced by a custom design with differential output, collecting the signal
both from the anode and the last dynode. The increased total resistance also reduced the heat production. Assuming a typical gain of $10^6$ and
a maximum hit rate of 5 MHz, 9 M$\Omega$ gives, at the operating voltage of 900V, a current that safely guarantees no gain drop during the operation of the PMT.
The signal is transported to the front-end board by a halogen-free shielded twinaxial cable, about 30 cm long, with 100 $\Omega$ impedance, equipped on both ends
with Harwin M80 connectors. The front-end board (fig. \ref{fig:NINOBoard}) consists of a motherboard with 64 analog differential inputs, 64 differential outputs, 
1 Embedded Local Monitor Board (ELMB) \cite{elmb} for services and remote control, and 8 mezzanines, with 1 NINO ASIC each.

Although the input impedance of the NINO mezzanine has been optimized to minimize reflections, the quality of the connection still leaves a residual reflection,
with a relative amplitude of a few percent, which, given the length of the cable, is visible at the end of the main signal, with about 4 ns delay. This results
in a double peak in the distribution of the time over threshold, where the secondary peak, at higher value, corresponds to signals large enough to have the reflection
above threshold. The ratio between the two peaks is a qualitative indicator of the position of the threshold with respect to the signal spectrum. Photon rates are such
to impose the single photoelectron regime on all PMTs, thus the efficiency of each PMT is defined by the fraction above threshold of its own single photoelectron spectrum (SER).
The gains of the PMTs vary by a factor of 10 and the
dependence of the PMT gain on the supply voltage is not sufficient to allow equalization of the
response of the PMTs. In order
to render the single channel inefficiency negligible the threshold should be low enough to cope with PMTs with gain as low as $5 \times 10^5$. This, together with the time performance
requirements, puts severe constraints on the acceptable level of noise.
The NINO threshold is set by a bias voltage that translates into an equivalent charge via the calibration factor 4 mV/fC. The resistive network on the NINO mezzanine allows
a minimum voltage of about 95 mV. Depending on the quality of the setup, a minimum threshold around 100 mV was proven to be achievable and set as baseline target,
guaranteeing a maximum inefficiency of a few percent on the PMTs with lowest gain.

The time resolution is a crucial parameter in NA62, as it is a high rate experiment, sensitive to accidentals. The intrinsic transit time spread of Hamamatsu
metal package PMTs, such as R7400-U03 and similar models, is about 300 ps. The contribution of the front-end has been kept below 100 ps by
the choice of the NINO ASIC (60 ps) and by limiting the noise; tests showed that noise below 30 fC gives a negligible contribution to the time jitter.
The cross talk has been measured to be below 1\% in the worst case.

\subsection{Read-out electronics}
\label{ssec:RO}
Each front-end board provides 64 LVDS outputs, distributed on 2 SCSI cables. Such signals are fed to the input of a 128 channel TDC board (TDCB) \cite{TDAQ}.
The TDCB is specifically designed for NA62, using 4 HPTDCs and one ALTERA Cyclone III FPGA. Each TEL62 board can house 4 TCDB
daughter boards giving a total of 512 channels. The HPTDC can measure both leading and trailing edges of the incoming signal with a dead time of 5 ns and 
a LSB of 100 ps. Although the signals from the chosen PMTs are shorter than 5 ns, it is still possible to measure both edges, and therefore the time over threshold, since
the NINO implements a stretching time of about 11 ns, which is added at the generation of digital output signals. The most severe limitation of the HPTDC, for the current application,
is the rate capability per channel: in order to keep the detection inefficiency below a few percent it has to be operated with an 80 MHz clock with a hit rate below 5 MHz.
This is true only for one single channel; since the internal buffers are shared by groups of 8 channels, also the rate capability is shared. Another limitation
is given by the output bandwidth of the TDCB to the TEL62, which is about 30 MHz of data words per HPTDC.
The hit rate per PMT is expected to range between 0.5MHz and 5MHz and therefore
the total rate per HPTDC is managed by distributing the channels using splitter boards, to use only one channel 
per each group of 8 of each HPTDC, with a
mapping based on MC estimates. This mapping keeps the rate per HPTDC below 15MHz which
corresponds to 30MHz of data words coming from the measurements of both the leading and
trailing edges.

\section{Technical run}
\begin{figure}[!t]
\centering
\includegraphics[width=0.25\textwidth]{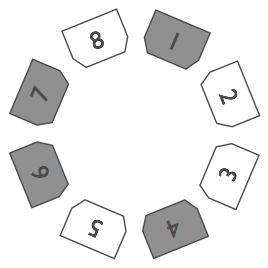}
\caption{Octants instrumented (grey) during the NA62 technical run}
\label{fig:KTAG2012}
\end{figure}

\begin{figure}[!b]
\centering
\includegraphics[width=0.5\textwidth]{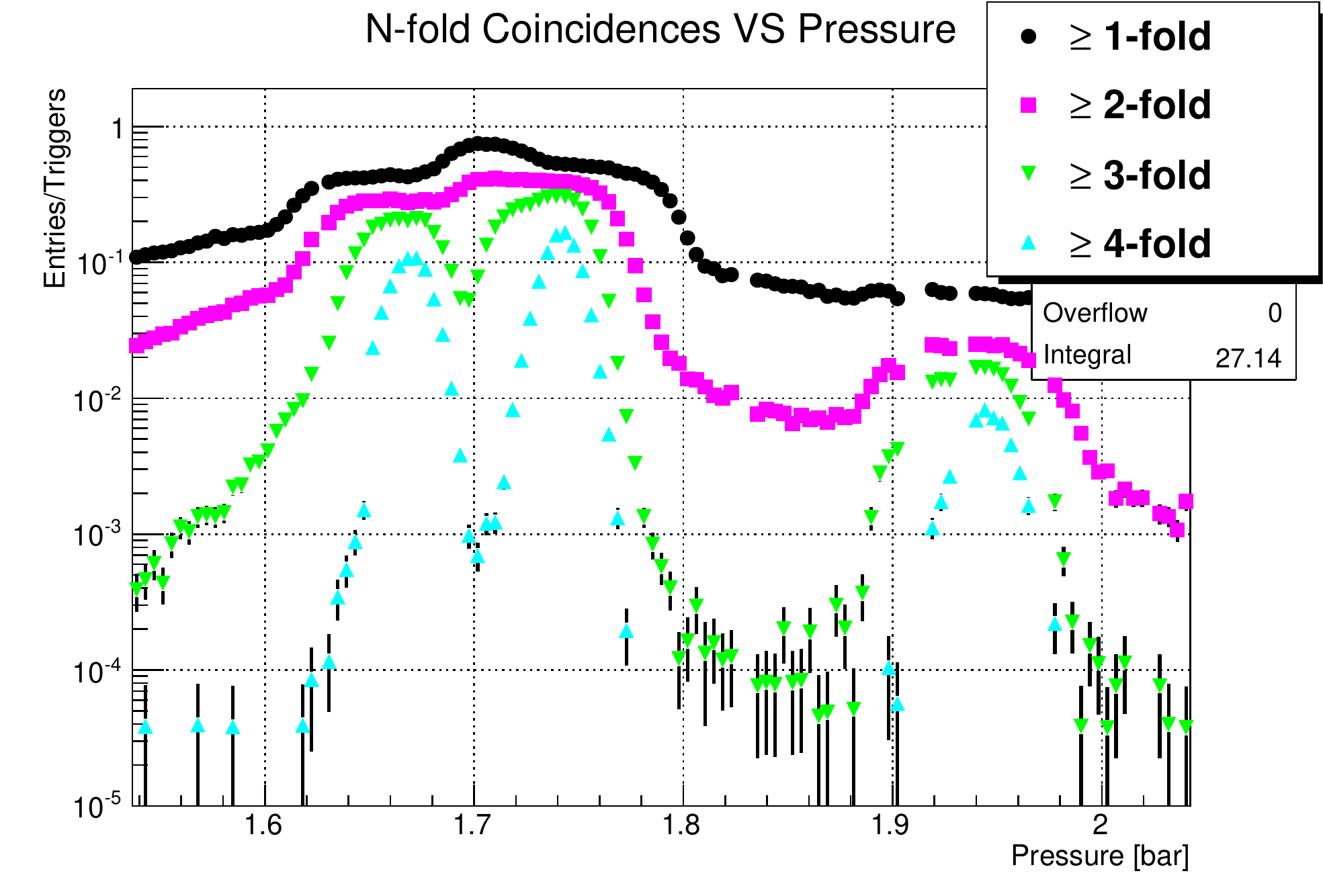}
\caption{Pressure scan performed during the NA62 technical run.}
\label{fig:PScan2012}
\end{figure}

\begin{figure}[!t]
\centering
\includegraphics[width=0.5\textwidth]{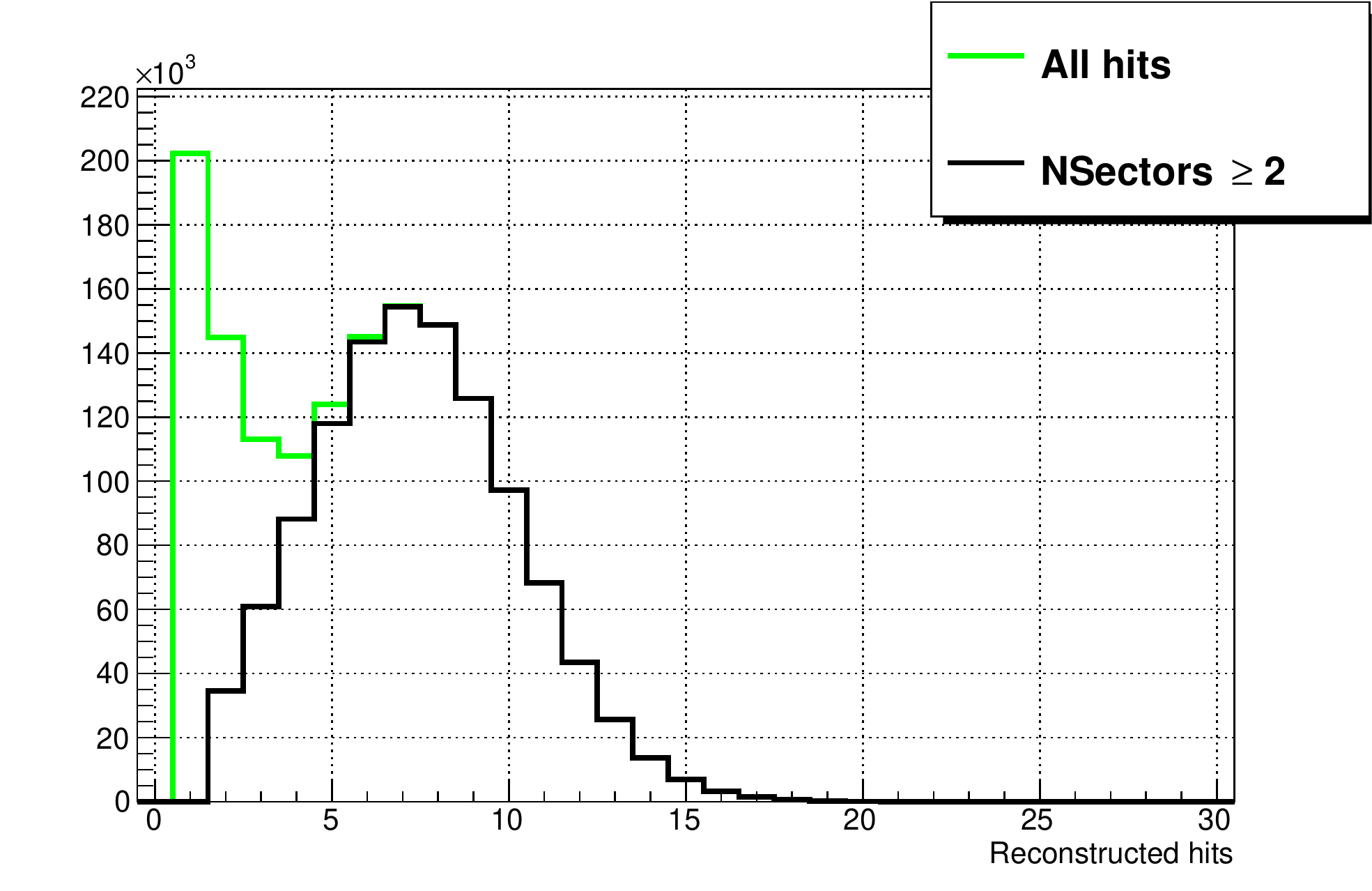}
\caption{Number of hits per beam particle. Accidentals are removed by requiring the coincidence between at least 2 octants.}
\label{fig:NHits2012}
\end{figure}

\begin{figure}[!b]
\centering
\includegraphics[width=0.5\textwidth]{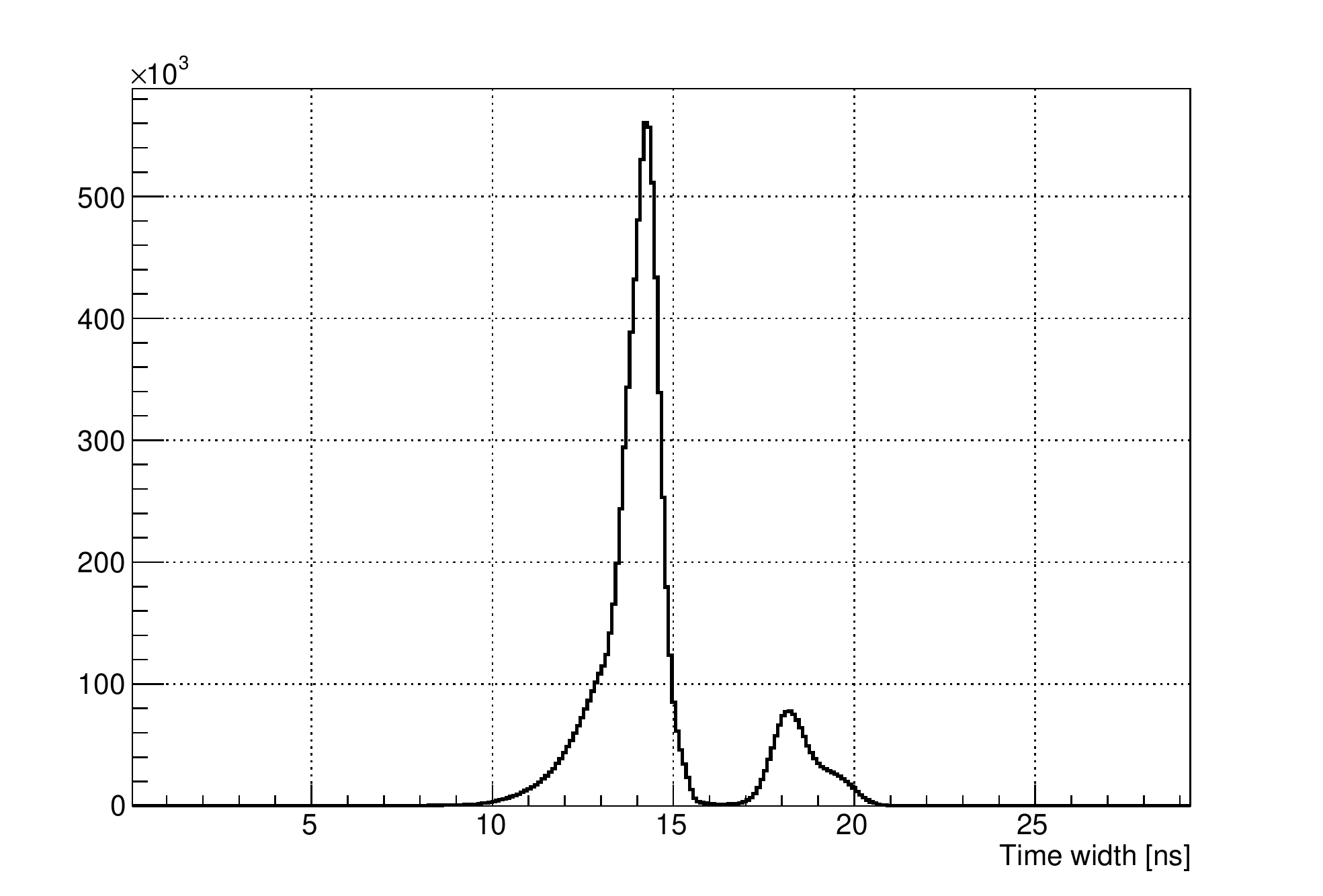}
\caption{Distribution of the time over threshold; the second peak is populated by events where a signal reflection is above threshold.}
\label{fig:ToT2012}
\end{figure}

In the autumn of 2012 NA62 took a sample of data with a partially instrumented detector, providing an opportunity to complete or refine the development of several components of the experiment. 
The mechanics of the KTAG were complete, but only 4 octants were equipped with LBs (fig. \ref{fig:KTAG2012}), and each LB had 32 PMTs installed. The beam intensity in the technical run was only 1-2\% of the nominal intensity and
therefore splitter boards were not required.
In order to keep the noise level within the foreseen limit, low pass filters had to be installed on the HV distribution boards. The prototype
front-end was not ready for remote control, therefore thresholds were set to a standard value of 270mV and could not be changed during the data taking, introducing a significant single channel inefficiency (up to 20\%).

The tuning procedure described in section \ref{ssec:TB2011Det} was adapted to the new setup and performed successfully (fig. \ref{fig:PScan2012}), although complicated by 
having only 4 instrumented octants. 
Fig. \ref{fig:NHits2012} shows the the number of hits per beam particle, with a clear separation between signal and background.

The partial setup of NA62 provided the means to perform a basic physics selection of the decay $K^+ \rightarrow \pi^+ \pi^0$ to perform efficiency studies with a kaon sample.

%\begin{figure}[!t]
%\centering
%\includegraphics[width=0.4\textwidth]{T02012}
%\caption{Time offset alignment}
%\label{fig:T02012}
%\end{figure}

\subsection{Time resolution}
KTAG has enough degrees of freedom to estimate its own time resolution without relying on an external time reference: the number of hits per kaon is large enough to
evaluate the time of the event by performing an average. The residuals with respect to the average are a measurement of the time resolution of an individual PMT, while
the global time resolution can be estimated by dividing by the square root of the number of hits. It is a slightly biased estimate, because of the correlation between average and residuals, 
but for a number of hits of the order of 10 or more the bias is negligible.
%\begin{figure}[!b]
%\centering
%\includegraphics[width=0.4\textwidth]{Slew2012}
%\caption{Time slewing correction}
%\label{fig:Slew2012}
%\end{figure}

\begin{figure}[!t]
\centering
\includegraphics[width=0.5\textwidth]{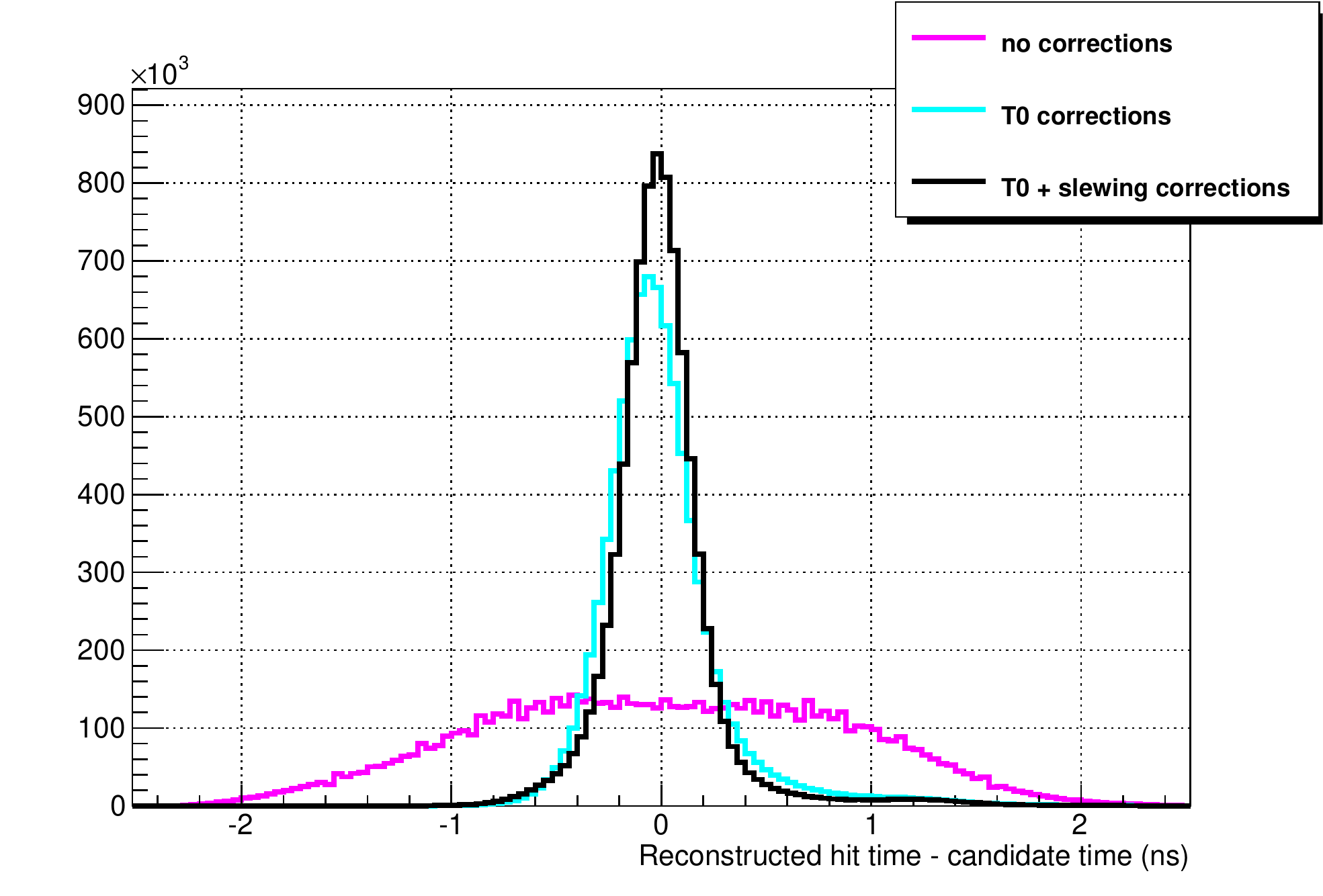}
\caption{Time resolution}
\label{fig:KTAGTRes2012}
\end{figure}

In order to achieve the best time performance, two sets of parameters must be evaluated: time offsets and time slewing parameters, exploiting the time over threshold (fig. \ref{fig:ToT2012}). 
Once these corrections are applied, the time resolution of the
individual PMT can be estimated (fig. \ref{fig:KTAGTRes2012}). For the partial setup the global time resolution is about 100 ps, which can be scaled to the final setup obtaining
60--70 ps.

\begin{figure}[!b]
\centering
\includegraphics[width=0.5\textwidth]{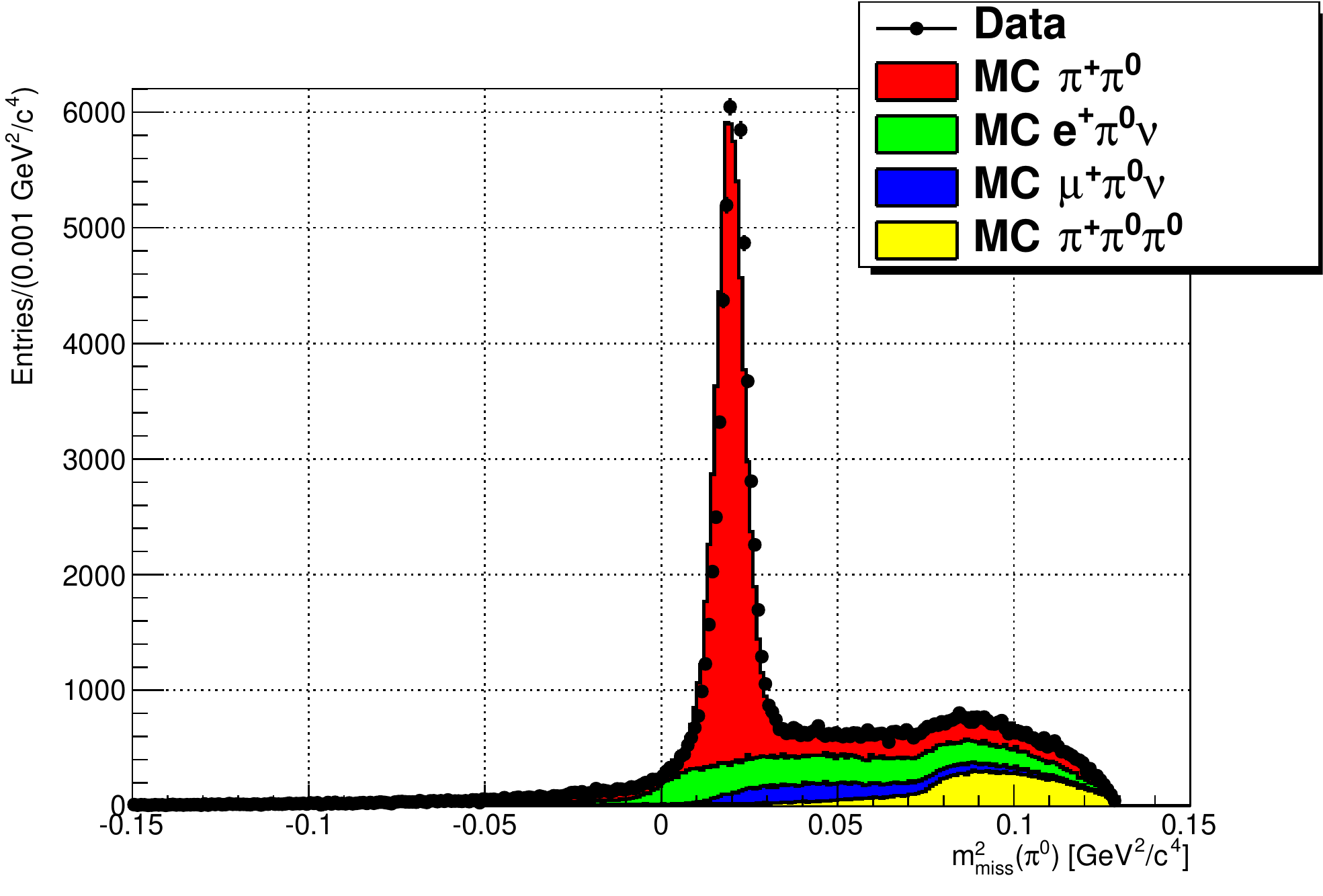}
\caption{Squared missing mass~$m_{miss}^2(\pi^0)$ obtained from the Technical Run, compared with the MC simulation of the most common $K^+$ decays involving at least a $\pi^0$. The MC samples are normalized according to their branching fractions.}
\label{fig:PiPi0}
\end{figure}

\subsection{Efficiency}
The presence of the Liquid Krypton calorimeter (LKr) in the partial setup, adding the kinematical constraint given by the beam geometry, allows a selection
of $\pi^0$s, which correspond only to kaon decays, mainly $K^+ \rightarrow \pi^+ \pi^0$ (fig. \ref{fig:PiPi0}). In this way a sample of kaons was isolated
during the data analysis, to be used as a reference for efficiency studies.
\begin{figure}[!t]
\centering
\includegraphics[width=0.5\textwidth]{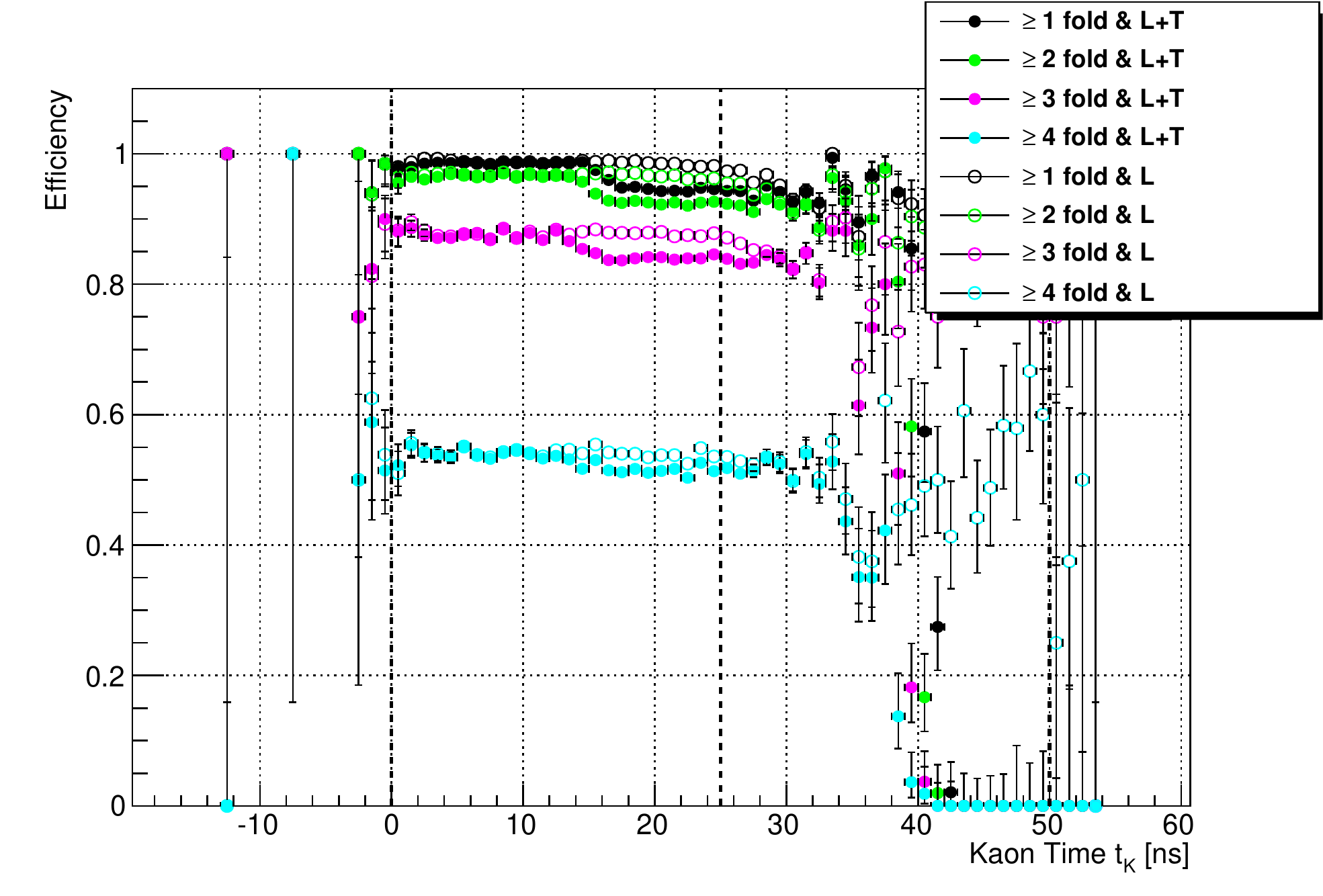}
\caption{CEDAR efficiency measured for the selected kaon sample as a function of the kaon decay time with respect to the trigger timestamp.
Two different edge requirements are considered: leading and trailing edge~(L+T); at least one leading~(L).}
\label{fig:KTAGEff2012}
\end{figure}

Fig. \ref{fig:KTAGEff2012} shows the efficiency as a function of the event time with respect to an external time reference; the drop in the second half of the
read-out window is due to the loss of trailing edges on its right hand side. The observation of this effect led to a modification of the TEL62 firmware to enlarge the read-out window and cope with fluctuations of the event position with respect to the trigger.

\section{Conclusions}
The performance of the CERN CEDAR was measured in a dedicated test beam in 2011, both in its original form and with the replacement of one PMT with a group of 3 Hamamatsu R7400 PMTs, new electronics, 
and a specially designed lightguide. This enabled a refinement of the electronics design and a Monte Carlo simulation of the Cherenkov photons to determine the most appropriate optical parameters. 
This information was incorporated into the design of KTAG, with a preliminary version, equipped with 4 octants each of 32 PMTs, participating in the NA62 technical run in Autumn 2012.
The information collected was used to finalize the design and estimate the final performance, taking into account
all the foreseen constraints.
The use of a preamplifier to match the PMT's analog output to the NINO input was discarded, and replaced with a custom voltage divider with differential output.
The Monte Carlo simulation was improved, leading to predictive results in the study of the final working parameters of the detector.
The light collection system, the services, the front-end and read-out electronics were built for the 2012 run, during which all design elements were evaluated, leading
to a final optimization.
Results from the 2012 data taking show that the final time resolution will be better than 100 ps, and that the required efficiency and pion rejection are achievable.
%\begin{figure}[!t]
%\begin{minipage}{0.46\textwidth}
%\centering
%\includegraphics[width=0.5\textwidth]{NPMTs}
%\caption{Number of hit PMTs at the beginning and at the end of the selection.}
%\label{fig:NPM}
%\end{minipage}\hfill
%\begin{minipage}{0.46\textwidth}
%\centering
%\includegraphics[width=0.5\textwidth]{BeamCompositionBest}
%\caption{Comparison between data and MC related to Cherenkov angle resolution and track angular resolution; differences in tails are
%probably due to beam halo or large angle scattering upstream the detector.}
%\label{fig:RingRadius}
%\end{minipage}\hfill
%\end{figure}

%\begin{figure}[!t]
%\begin{minipage}{0.46\textwidth}
%%\centering
%%\includegraphics[width=0.5\textwidth]{RingRadiusTLessT}
%%\caption{Comparison between reconstructed Cherenkov radius in two single burst data, close in time, acquired with and without external
%%trigger from scintillators. No normalization is applied.}
%%\label{fig:RingRadiusTLessT}
%\centering
%\includegraphics[width=0.5\textwidth]{BeamCompositionBest}
%\caption{Comparison between data and MC related to Cherenkov angle resolution and track angular resolution; differences in tails are
%probably due to beam halo or large angle scattering upstream the detector.}
%\label{fig:RingRadius}
%\end{minipage}\hfill
%\begin{minipage}{0.46\textwidth}
%\centering
%\includegraphics[width=0.5\textwidth]{TrigWinBreakDown}
%\caption{Decomposition of raw hits time distribution based on the multiplicity per $1ns$ bin}
%\label{fig:TrigWinBreakDown}
%\end{minipage}\hfill
%\end{figure}

\section*{Acknowledgments}
For the test beam in 2011 we would like to thank CERN and the beam group, in particular J. Spanggaard and E. Gschwendtner, for their
support in setting up and handling the beamline and the original read-out and control of the CEDAR; we are also grateful to the NA62
RICH working group for providing us with their prototype front-end electronics.
In 2012 the support of NA62, for the central data acquisition system, N. Doble and L. Gatignon, for the K12 beam line, was instrumental
in exploiting the technical run for testing the performance of the CEDAR.
We would like to thank Thomas Schneider for his help in aluminising lenses to produce mirrors of high reflectivity, and Crispin Williams for providing the NINO mezzanine cards together with advice and assistance in their usage.
We would like to thank the staff of Liverpool University workshop for their important contribution to this work.
It is a pleasure to acknowledge the Science and Technology Facility Council UK, the Royal Society and the European Research Council
for their generous funding of this project.
%\appendix

%\section*{Acknowledgments}
%
%Acknowledgments should be inserted at the end of the paper, 
%before the references, not as a footnote to the title. 
%Use an unnumbered section heading for the Acknowledgments.

%
%
%===============================================================
%
%   References
%
%===============================================================
%
%

\end{document}